\documentstyle[12pt]{article}

\begin{document}

\begin{flushright}
IMSc/2013/04/03 
\end{flushright} 

\vspace{2mm}

\vspace{2ex}

\begin{center}

{\large \bf M theory Branes : U duality properties \\
\vspace{2ex} 
and a class of new Static Solutions } \\

\vspace{8ex}

{\large  S. Kalyana Rama}

\vspace{3ex}

Institute of Mathematical Sciences, C. I. T. Campus, 

Tharamani, CHENNAI 600 113, India. 

\vspace{1ex}

email: krama@imsc.res.in \\ 

\end{center}

\vspace{6ex}

\centerline{ABSTRACT}

\begin{quote} 

We obtain the most general static intersecting brane solutions
by directly solving the relevant equations of motion
analytically and in complete generality. These solutions reduce
to the known ones in special cases, and contain further a class
of new static solutions which are horizonless. We describe their
properties and discuss their physical relevance. Along the way,
we also describe the features of the brane energy momentum
tensors, the equations of motion, and their solutions which
arise as consequences of the intersection rules and the U
duality symmetries of M theory.

\end{quote}

\vspace{2ex}











\newpage

\vspace{4ex}

\centerline{\bf 1. Introduction}  

\vspace{2ex}

Consider a static spherically symmetric star. If it is
sufficiently massive then it is likely to collapse and to
finally form a black hole or, possibly, some other static
object. It is self evident that, generically, the fields outside
such stars and the final collapsed objects must be described by
the most general static solutions to the relevant equations of
motion. In Einstein's general theory of relativity, for example,
the most general static spherically symmetric solution is given
by the Schwarzschild solution. In Brans--Dicke theory, the most
general static spherically symmetric solutions are given by the
Janis--Newman--Winicour--Wyman (JNWW) solutions \cite{jnw}, the
Schwarzschild solution now being a special case.

In M theory (or equivalently string theory), the black holes are
described by various intersecting brane configurations where the
branes wrap the compact internal spaces, taken to be toroidal.
For example, the four dimensional black holes are described in M
theory by two stacks of $M2$ branes and two stacks of $M5$
branes which wrap the seven dimensional compact toroidal space
and intersect according to the BPS rules. \footnote{ According
to the BPS rules, two stacks of five branes intersect along
three common spatial directions; a stack each of two branes and
five branes intersect along one common spatial direction; and
two stacks of two branes intersect along zero common spatial
direction. Also, the branes are taken to be uniformly smeared
along the remaining compact directions \cite{t96, alg}.} The
standard intersecting brane solutions are given in \cite{t96,
alg, hs, hm, rt97, aeh}. They are static solutions independent
of compact toroidal coordinates and are spherically symmetric in
the non compact transverse space.

It is natural to expect that such black holes in M theory must
have been formed by the collapse of sufficiently massive stars.
In order to study such collapses, one first needs to construct a
static star in M theory. This requires, among other things, a
knowledge of the most general static solutions to the relevant
equations of motion.

The standard brane solutions given in \cite{t96, alg, hs, hm,
rt97, aeh} are not the most general ones. These solutions can be
thought of as analogous to the Schwarzschild solutions rather
than to the JNWW solutions. Brane solutions of the JNWW type
also exist. They have been found by several groups in different
forms and using different methods \cite{zz, k05}, see \cite{ar}
also. However, in all these works, while solving the equations
of motion, some ansatz or the other is made regarding the form
of the brane solutions which thereby limits their generality.
Therefore, these solutions are not the most general ones.

In this paper, we find the most general static intersecting
brane solutions. We directly solve the relevant equations of
motion analytically and in complete generality. No ansatz is
made regarding the form of the solutions. The resulting
solutions are thus completely general. In special cases, they
reduce to the standard solutions of \cite{t96, alg, hs, hm,
rt97, aeh} and to the more general ones of \cite{zz, k05}. The
general solutions we find contain further a class of new static
solutions which are horizonless and, to the best of our
knowledge, have not appeared in the literature. \footnote{In our
earlier reports \cite{k11}, we obtained the most general vacuum
solutions first and then used it to generate the brane solutions
by suitable boosting and U duality operations \cite{rt97}. The
brane solutions thus generated turn out to be the same as the
ones obtained in this paper by directly solving the equations of
motion.}

We point out here that, except the Schwarzschild and the
standard brane solutions, the general solutions mentioned above
-- namely, JNWW solutions, those given in \cite{zz, k05}, and
the new ones found in this paper -- all have, generically, a
singularity either at a finite radius or at a vanishing radius.
These singularities are not covered by any horizon and, hence,
are naked. Also, they are present even though the solutions all
have positive ADM mass. In the case of a static star, the
interior solutions get modified and, hence, these singularities
are not relevant since the general solutions are applicable only
outside the star. The general solutions can then be used to
explore the observational consequences by studying the motions
of various probes outside such a static star, as in \cite{will,
pankaj, narayan, panda} for example.

Also, such general solutions are essential ingredients in the
study of collapse of stars, and in determining the nature and
the properties of the final collapsed objects; in particular, in
determining whether a naked singularity appears or not during
and after the collapse. In the context of four dimensional JNWW
solutions, such collapse processes and various issues involving
naked singularities have been extensively studied in
\cite{narayan, gj}. Similar studies are also needed in the
present context.

In this paper, we describe our general static intersecting brane
solutions. We start with the eleven dimensional low energy
effective action for the intersecting branes of M theory, and
write down the equations of motion. These equations, their
solutions, and the energy momentum tensors for the branes all
exhibit several features which arise as consequences of the BPS
intersection rules and the underlying U duality symmetries of M
theory. We describe these features along our way to solving the
equations of motion. We first solve the equations and obtain the
general analytical solutions in terms of a variable $\tau$ which
is suggested naturally by the equations themselves. We then
obtain the general relation between $\tau$ and the standard
radial coordinate $r$, which completes the solutions. All along,
we take particular care in ensuring that no generality is lost
at any stage. We then describe the properties of the resulting
solutions.

The paper is organised as follows. In section {\bf 2}, we
present the action and the metric which can describe the static
intersecting branes. In section {\bf 3}, we write down the
equations of motion in several convenient forms and discuss the
structure of the solutions. In section {\bf 4}, we describe the
features which are consequences of the intersection rules and
the U duality symmetries of M theory. In section {\bf 5}, we
solve the equations of motion in complete generality in terms of
$\tau \;$. In section {\bf 6}, we obtain the general relation
between $\tau$ and the radial coordinate $r \;$. In section {\bf
7}, we describe the properties of the solutions pointing out the
special cases which lead to the solutions known before, and the
general cases which lead to the new solutions. In section {\bf
8}, we make several remarks which illustrate various features of
the present solutions. In section {\bf 9}, we conclude with a
brief summary and by mentioning a few issues for further
studies.

Three Appendices contain some useful expressions and formulae.
In Appendix {\bf A}, we give the non vanishing components of the
Riemann tensor for the metric used here. In Appendix {\bf B},
following the analysis of \cite{jxlu}, we give the general
expressions for the ADM mass. In Appendix {\bf C}, for the sake
of completeness, we give the general vacuum solutions obtained
earlier \cite{k11} directly in terms of $r \;$.

\underline{A Note to the reader :} 
The paper is somewhat long and contains many technical details
which we have presented in order to solve the equations
analytically and, at the same time, to make the generality of
our approach obvious. Much of these details may be familiar to
the reader. Hence, for the sake of the reader's convenience, we
have collected the main results and the expressions for the
general solutions together in one place in the beginning of
section {\bf 7} before proceeding to discuss their properties in
sections {\bf 7} and {\bf 8}. So those readers who are mainly
interested in the solutions may go straight to these sections at
a first reading.


\vspace{4ex}

\centerline{\bf 2. General Set Up}  

\vspace{2ex}

We study the static solutions describing ${\cal N}$ stacks of M
theory branes, intersecting according to the BPS rules
\cite{t96, alg} whereby two stacks of five branes intersect
along three common spatial directions; a stack each of two
branes and five branes intersect along one common spatial
direction; and two stacks of two branes intersect along zero
common spatial direction. We start with the low energy effective
action for M theory branes and an appropriate static ansatz for
the metric.

The relevant part of the eleven dimensional low energy effective
action describing ${\cal N}$ stacks of M theory branes mentioned
above may be written in standard notation as
\begin{equation}\label{s0} 
S = \frac{1}{2 \kappa^2} \; \int d^{11} x \; \sqrt{- g} \; 
\left( {\cal R} - \sum_I \frac {F^2_{(I)}}
{2 (n_I + 2)!} \; \right)
\end{equation} 
where $I = 1, 2, \cdots, {\cal N} \;$, 
\[
F^2_{(I)} 
= \sum_{M_1 \cdots M_{n_I + 2}}  
F_{M_1 M_2 \cdots M_{n_I + 2}} \; 
F^{M_1 M_2 \cdots M_{n_I + 2}} \; \; , 
\] 
$F_{M_1 \cdots M_{n_I + 2}} \;$ is the $(n_I + 2)-$form field
strength for the $I^{th}$ stack of $n_I-$branes, and $\; n_I =
2$ or $5 \;$. We write the summation indices explicitly since
not all the sums in this paper are over repeated indices. The
above form of the brane action can be shown to follow as a
consequence of the BPS intersection rules, see \cite{aeh,
ar}. Setting $\kappa^2 = 1 \;$, the resulting equations of
motion are given by
\begin{eqnarray}
{\cal R}_{M N} - \frac{1}{2} \; g_{M N}\; {\cal R} & = &
T_{M N} \; = \; \sum_I T_{M N (I)} \label{rmn} \\
\sum_M \; \partial_M \left( \sqrt{- g} \; 
F^{M M_2 \cdots M_{n_I + 2}} \right) & = & 0 \label{dmfm} \\
\; \; \; \longleftrightarrow \; \; \; 
\sum_M \; \nabla_M \; T^M_{\; \; N (I)} & = & 0 \label{tmnI}
\end{eqnarray}
where $T_{M N}$ is the total energy momentum tensor and
\begin{equation}\label{bhtmnI}
T_{M N (I)} = \frac{1}{2 (n_I + 1) !}  \left( 
\sum_{M_2 \cdots M_{n_I + 2}}  
F_{M M_2 \cdots M_{n_I + 2}} \; 
F_N^{\; \; M_2 \cdots M_{n_I + 2}}
- \frac{g_{M N} \; F^2_{(I)}}{2 (n_I + 2)} \right)
\end{equation}
is the energy momentum tensor for the $I^{th}$ stack of branes.

We take the spatial directions of the brane worldvolumes to be
toroidal and assume necessary isometries. Let the spacetime
coordinates be given by $x^M = (t, x^i, r, \theta^a)$ where $i =
1, 2, \cdots, n_c$ and $a = 1, 2, \cdots, m \;$. The total
spacetime dimension $D = n_c + m + 2 = 11 \;$. The coordinates
$x^i$ describe the compact, $n_c$ dimensional, toroidal space;
and the radial coordinate $r$ and the spherical coordinates
$\theta^a$ describe the non compact, $(m + 1)$ dimensional,
transverse space. In the following, we will assume that $m \ge 2
\;$ and study the static solutions. The line element $d s$ which
can describe the static intersecting branes may now be written
as
\begin{equation}\label{ds} 
d s^2 = g_{M N} \; d x^M \; d x^N 
= - e^{2 \lambda^0} d t^2 + \sum_i e^{2 \lambda^i} (d x^i)^2
+ e^{2 \lambda} d r^2 + e^{2 \sigma} d \Omega_m^2
\end{equation}
where $d \Omega_m$ is the standard line element on an $m$
dimensional unit sphere. The non vanishing components of the
field strengths are $F_{0 i_1 \cdots i_{n_I} r} = \partial_r
{\cal A}_{(I)} \;$ where $i_1, \cdots, i_{n_I}$ and ${\cal
A}_{(I)}$ denote, respectively, the spatial worldvolume
directions and the $(n_I + 1)-$form gauge field for the $I^{th}$
stack of branes. The fields $(\lambda^0, \lambda^i, \lambda,
\sigma)$ and ${\cal A}_{(I)}$ are functions of $r \;$ only.

It can now be seen that the energy momentum tensors $T_{M N
(I)}$ given in equation (\ref{bhtmnI}) are all diagonal. We
denote these diagonal elements as 
\begin{equation}\label{rhopiI}
\left( T^0_{\; \; 0 (I)}, \; T^i_{\; \; i (I)}, \; 
T^r_{\; \; r (I)}, \; T^a_{\; \; a (I)} \right) = 
\left( - \rho_I, \; p_{i I}, \; \Pi_I, \; p_{a I} \right) 
\end{equation}
where $p_{a I} = p_{\Omega I}$ for all $a \;$. The total energy
momentum tensor is now given by $T^M_{\; \; \; \; N} = diag \;
\left( - \rho, \; p_i, \; \Pi, \; p_a \right)$ where
\begin{equation}\label{rhopitotal}
\rho = \sum_I \rho_I \; \; , \; \; \; 
p_i = \sum_I p_{i I} \; \; , \; \; \; 
\Pi = \sum_I \Pi_I \; \; , \; \; \; 
p_a = p_\Omega = \sum_I p_{\Omega I} \; \; .
\end{equation}
From the expression for $T_{M N (I)}$ in equation
(\ref{bhtmnI}), it follows that $p_{i I} = p_{\parallel I} \;$
or $p_{\perp I}$ according to whether the $x^i$ direction is
parallel or transverse to the worldvolume of the $I^{th}$ stack
of branes, and further that
\begin{equation}\label{rhopianswer}
- \rho_I = p_{\parallel I} = - p_{\perp I} = \Pi_I 
= - p_{\Omega I} = \frac{1}{4} \; F_{0 i_1 \cdots i_{n_I} r} \;
F^{0 i_1 \cdots i_{n_I} r} \; \; .
\end{equation}

Note that, in the studies of stars or of cosmological evolution,
one often assumes that the total energy momentum tensor is the
sum of ${\cal N}$ individual energy momentum tensors each of
which is diagonal and seperately conserved; and then supplements
the equations of motion with equations of state for each
individual $I$ which determine the diagonal components $T^M_{\;
\; M (I)}$ in terms of $\rho_I$, or equivalently $\Pi_I \;$. In
our case here, these properties of the energy momentum tensors
follow from an underlying action for branes and from the BPS
intersection rules, and equation (\ref{rhopianswer}) provides
the equations of state for each individual $I \;$.


\vspace{4ex}

\centerline{\bf 3. Equations of motion}  

\vspace{2ex}

We first write down the equations of motion for the fields
$\Pi_I \;$ and $(\lambda^0, \lambda^i, \lambda, \sigma)$ using
equations (\ref{rmn}), (\ref{tmnI}), (\ref{ds}), (\ref{rhopiI}),
and (\ref{rhopitotal}) only. In particular, we will not use
equations (\ref{bhtmnI}) or (\ref{rhopianswer}). The resulting
equations are, therefore, valid quite generally wherever the
metric is given by equation (\ref{ds}) and where the total
energy momentum tensor is the sum of ${\cal N}$ energy momentum
tensors each of which is diagonal and seperately conserved.

Let
\begin{equation}\label{palphas} 
\alpha = (0, i, a) 
\; \; , \; \; \; 
\lambda^\alpha = (\lambda^0, \lambda^i, \lambda^a) 
\; \; , \; \; \; 
p_{\alpha I} = (p_{0 I}, p_{i I}, p_{a I}) 
\end{equation}
where $p_{0 I} = - \rho_I \;$, and $\lambda^a = \sigma$ and
$p_{a I} = p_{\Omega I}$ for all $a \;$. Define
\begin{eqnarray}
\Lambda & = & \sum_\alpha \lambda^\alpha = 
\lambda^0 + \sum_i \lambda^i + m \sigma \label{Lambda} \\
T_I & = & \sum_M T^M_{\; \; \; M (I)} = 
\Pi_I + \sum_\alpha p_{\alpha I} \; \; . \label{T}
\end{eqnarray}
It is straighforward to calculate the Riemann tensor
corresponding to the metric given by equation (\ref{ds}). We
have listed its non vanishing components in Appendix {\bf
A}. Using these components and the above definitions, it follows
after some algebra that equations (\ref{rmn}), (\ref{tmnI}),
(\ref{ds}), (\ref{rhopiI}), and (\ref{rhopitotal}) give
\begin{eqnarray}
(\Pi_I)_r & = & - \Pi_I \; \Lambda_r 
+ \sum_\alpha p_{\alpha I} \lambda^\alpha_r \label{Pir} \\
\Lambda^2_r - \sum_\alpha (\lambda^\alpha_r)^2 & = &
2 \sum_I \Pi_I \; e^{2 \lambda} 
+ m (m - 1) \; e^{2 \lambda - 2 \sigma} \label{Lambdar2}
\end{eqnarray}
\begin{equation}\label{alpharr}
\lambda^\alpha_{r r} + (\Lambda_r - \lambda_r) \;
\lambda^\alpha_r = \sum_I \left(- p_{\alpha I} 
+ \frac{T_I}{D - 2} \right) \; e^{2 \lambda} 
+ \delta^{\alpha a} \; (m - 1) \; e^{2 \lambda - 2 \sigma}
\end{equation}
where the subscripts $r$ denote $r-$derivatives. 

Equation (\ref{alpharr}) now suggests a change of variable from
$r$ to $\tau$ given by \footnote{ We note in passing that this
variable $\tau$ and the solutions obtained in terms of $\tau$
bear a striking similarity to those which appear in the `H-FGK
formalism' \cite{fgk}. Upon substitution of the solution for
$\chi (\tau)$, see later in the following, the form of the $(m +
1)$ dimensional line element given in equation (\ref{dsm+1})
becomes an example of such a similarity; the $Mp$ brane
solutions given in equation (\ref{mptau}) below is another
example.}
\begin{equation}\label{rtau}
e^\lambda \; d r = e^\Lambda \; d \tau \; \; . 
\end{equation}
Then, for any function $X(r(\tau))$, we have
\begin{equation}\label{Xtaur}
X_\tau = e^{\Lambda - \lambda} \; X_r
\; \; , \; \; \; 
X_{\tau \tau} = e^{2 (\Lambda - \lambda)} \; \left( X_{r r} 
+ (\Lambda_r - \lambda_r) \; X_r \right)
\end{equation}
where the subscripts $\tau$ denote $\tau-$derivatives. The line
element for the $(m + 1)$ dimensional transverse space given in
equation (\ref{ds}) now becomes
\begin{equation}\label{dsm+1} 
e^{2 \lambda} d r^2 + e^{2 \sigma} d \Omega_m^2 \; = \; 
e^{2 \sigma} \; \left( e^{2 \chi} \; d \tau^2 + d \Omega_m^2
\right) 
\end{equation}
where the field $\chi$ is defined by 
\begin{equation}\label{chidefn}
\chi = \Lambda - \sigma = 
\lambda^0 + \sum_i \lambda^i + (m - 1) \sigma \; \; . 
\end{equation}
Equations (\ref{Pir}) -- (\ref{alpharr}), expressed in terms of
$\tau$, become
\begin{eqnarray}
(\Pi_I)_\tau & = & - \Pi_I \; \Lambda_\tau 
+ \sum_\alpha p_{\alpha I} \lambda^\alpha_\tau \label{Pitau} \\
\Lambda^2_\tau - \sum_\alpha (\lambda^\alpha_\tau)^2 & = &
2 \sum_I \Pi_I \; e^{2 \Lambda} 
+ m (m - 1) \; e^{2 \chi} \label{Lambdatau2} \\
\lambda^\alpha_{\tau \tau} 
& = & \sum_I \left(- p_{\alpha I} + \frac{T_I}{D - 2} \right) \;
e^{2 \Lambda} + \delta^{\alpha a} \; (m - 1) \; e^{2 \chi}
\; \; . \label{alphatautau}
\end{eqnarray}
Equation for the field $\chi$ defined in equation
(\ref{chidefn}) can now be obtained. It is given by
\begin{equation}\label{L-sigmatautau}
\chi_{\tau \tau} = \sum_I \left( \Pi_I + p_{\Omega I} \right) \;
e^{2 \Lambda} + (m - 1)^2 \; e^{2 \chi} \; \; .
\end{equation}
The above equations may be written more compactly upon defining
$G_{\alpha \beta}$ and $G^{\alpha \beta}$ by
\begin{equation}\label{Galphabeta}
G_{\alpha \beta} = 1 - \delta_{\alpha \beta} 
\; \; , \; \; \; 
G^{\alpha \beta} = \frac{1}{D - 2} - \delta^{\alpha \beta} 
\; \; ,
\end{equation}
and upon defining $h_{\alpha I}$ by 
\begin{equation}\label{halpha}
h_{\alpha I} = \sum_\beta G_{\alpha \beta} \left(- p_{\beta I} 
+ \frac{T_I}{D - 2} \right) = \Pi_I + p_{\alpha I} \; \; .
\end{equation}
Then we get 
\begin{eqnarray}
(\Pi_I)_\tau & = & - 2 \Pi_I \; \Lambda_\tau 
+ \sum_\alpha h_{\alpha I} \lambda^\alpha_\tau \label{2Pitau} \\
\sum_{\alpha \beta} G_{\alpha \beta} \; \lambda^\alpha_\tau \;
\lambda^\beta_\tau & = & 2 \sum_I \Pi_I \; e^{2 \Lambda} 
+ m (m - 1) \; e^{2 \chi} \label{2Lambdatau2} \\
\lambda^\alpha_{\tau \tau} & = & 
\sum_{\beta I} G^{\alpha \beta} \; h_{\beta I} \; 
e^{2 \Lambda} + \delta^{\alpha a} \; (m - 1) \; 
e^{2 \chi} \; \; . \label{2alphatautau} \\
\chi_{\tau \tau} & = & \sum_I h_{\Omega I} \; e^{2 \Lambda} 
+ (m - 1)^2 \; e^{2 \chi} \; \; .  \label{2L-sigmatautau}
\end{eqnarray}

We require the solutions to correspond to asymptotically flat
spacetime in the limit $r \to \infty \;$. Therefore, the above
equations of motion are to be solved with the asymptotic
condition that, in the limit $r \to \infty$, the fields
\begin{equation}\label{bc}
\left( e^{\lambda^0}, \; e^{\lambda^i}, \; e^\lambda, \; 
e^\sigma \; ; \; \; {\cal A}_{(I)} \right)
\; \to \; \left( 1, \; 1, \; 1, \; r \; ; \; \;
\frac{Q_I}{r^{m - 1}} \right)
\end{equation}
where $Q_I$ are the charges. The definitions of $(\Lambda,
\chi)$ and equation (\ref{rtau}) then imply that, in the limit
$r \to \infty \;$,
\begin{equation}\label{bc2}
e^\Lambda \to r^m \; \; , \; \; \;
e^\chi \to r^{m - 1} \; \; , \; \; \;
(\tau_0 - \tau) \; \to \; \frac {1} {(m - 1) \; r^{m - 1}} 
\end{equation}
where $\tau_0$ is a constant. Thus $\tau \to \tau_0$ from below
in the limit $r \to \infty \;$. Furthermore, since $\frac{d r}{d
\tau}$ is positive, it follows that $\tau$ decreases from
$\tau_0$ as $r$ decreases from $\infty \;$. Hence $\tau \le
\tau_0$ for $r \le \infty$ and, incorporating this feature,
equation (\ref{rtau}) may be written as
\[
\int^{\tau_0}_\tau d \tau = 
\int^\infty_r d r \; e^{\lambda - \Lambda} \; \; . 
\]
The constant $\tau_0$ can be set to zero with no loss of
generality. But we will not do so in the following in order to
avoid the seemingly spurious negative signs in expressions
involving $(\tau_0 - \tau) \;$.

Note that, for a metric of the form given in equation (\ref{ds})
and which corresponds to asymptotically flat spacetime, one can
obtain the ADM mass. In Appendix {\bf B}, following the analysis
of \cite{jxlu}, we have given the corresponding general
expressions for the ADM mass.


\vspace{4ex}

\centerline{\bf Linear equations of state}  

\vspace{2ex}

To solve equations (\ref{2Pitau}) -- (\ref{2L-sigmatautau}), one
requires the equations of state which give, for example,
$p_{\alpha I} = (p_{0 I}, p_{i I}, p_{a I})$ as functions of
$\Pi_I \;$. Consider the case where these equations of state are
linear and are given by
\begin{equation}\label{eos}
h_{\alpha I} = \Pi_I + p_{\alpha I} = z^I_\alpha \; \Pi_I
\end{equation}
where $z^I_\alpha$ are constants and $z^I_a = z^I_\Omega$ for
all $a \;$. Equation (\ref{2Pitau}) can now be solved to give
\begin{equation}\label{PiIhI} 
\Pi_I = \Pi_{I 0} \; e^{- 2 \Lambda + l^I} 
\; \; \; \longrightarrow \; \; \; 
h_{\alpha I} \; e^{2 \Lambda} = z^I_\alpha \; 
\left( \Pi_{I 0} \; e^{l^I} \right)
\end{equation}
where $\Pi_{I 0}$ are constants and $l^I$ are given by
\begin{equation}\label{lIdefn} 
l^I = \sum_\alpha z^I_\alpha \; \lambda^\alpha \; \; . 
\end{equation}
Using these expressions, equations for $\lambda^\alpha$, $\;
l^I$, and $\chi$ become
\begin{eqnarray}
\lambda^\alpha_{\tau \tau} & = & \sum_I z^{\alpha I} \; 
\left( \Pi_{I 0} \; e^{l^I} \right)
+ \delta^{\alpha a} \; (m - 1) \; e^{2 \chi} \label{lalphatt} \\
l^I_{\tau \tau} & = & \sum_J {\cal G}^{I J} \; 
\left( \Pi_{J 0} \; e^{l^J} \right) 
+ z^I_\Omega \; m (m - 1) \; e^{2 \chi} \label{lItt} \\
\chi_{\tau \tau} & = & \sum_I z^I_\Omega \; 
\left( \Pi_{I 0} \; e^{l^I} \right)
+ (m - 1)^2 \; e^{2 \chi}  \label{chitt}
\end{eqnarray}
where 
\begin{equation}\label{GIJ}
z^{\alpha I} =  \sum_\beta G^{\alpha \beta} \; z^I_\beta
\; \;, \; \; \; 
{\cal G}^{I J} = \sum_{\alpha \beta} 
G^{\alpha \beta} z^I_\alpha \; z^J_\beta \; \; . 
\end{equation}

One may now express $\lambda^\alpha$ in terms of $l^I$ and
$\chi$ as follows. Solve equations (\ref{lItt}) and
(\ref{chitt}) as simultaneous equations to obtain $\Pi_{I 0} \;
e^{l^I}$ and $e^{2 \chi}$ in terms of $l^I_{\tau \tau}$ and
$\chi_{\tau \tau} \;$. Substituting these expressions into
equation (\ref{lalphatt}) gives an equation of the form
\[
\lambda^\alpha_{\tau \tau} = X^\alpha_{\tau \tau}
\; \; , \; \; \; 
X^\alpha = \sum_I a^\alpha_I \; l^I + a'^\alpha \chi
\]
where $a^\alpha_I$ and $a'^\alpha$ are constants. It then
follows that the most general solution for $\lambda^\alpha$ is
given by $\lambda^\alpha = X^\alpha + L^\alpha \tau + a''^\alpha
\;$ where $L^\alpha$ and $a''^\alpha$ are integration constants.
Substituting this expression for $\lambda^\alpha$ into the
definitions $l^I = \sum_\alpha z^I_\alpha \lambda^\alpha$ and
$\chi = \Lambda - \sigma$ will lead to ${\cal N} + 1$
consistency constraints each on $L^\alpha$ and $a''^\alpha \;$.
A complete solution thus follows once equations (\ref{lItt}) and
(\ref{chitt}) are solved for $l^I (\tau)$ and $\chi (\tau)
\;$. Finally, equation (\ref{2Lambdatau2}) will lead to one
further constraint among various integration constants appearing
in the complete solutions for $(l^I, \; \chi, \; \lambda^\alpha)
\;$.

As pointed out earlier, the various forms of equations of motion
above have been obtained without using the expression for $T_{M
N (I)}$ given in equation (\ref{bhtmnI}) or (\ref{rhopianswer}).
Hence, these equations of motion are all valid quite generally
wherever the metric is given by equation (\ref{ds}) and the
total energy momentum tensor is the sum of ${\cal N}$ energy
momentum tensors, each of which is diagonal and seperately
conserved. Thus, for example, these equations may also be used
to study stars made up of M theory branes; similarly, with a few
straightforward modifications involving $t$ and $r$, and
$\rho_I$ and $\Pi_I$, they may also be used to study the
cosmological evolution of a universe made up of intersecting M
theory branes \cite{cm, k07, bdk, bk, b}.


\vspace{4ex}

\centerline{\bf 4. M theory branes}

\vspace{2ex}

Whether or not the equations for $l^I (\tau)$ and $\chi (\tau)
\;$ can be solved explicitly depends on the values of the
constants $z^I_\alpha \;$ and the consequent structure of ${\cal
G}^{I J} \;$. It does not seem to be possible to obtain explicit
solutions for general arbitrary values of $z^I_\alpha \;$. Hence
we now specialise to M theory branes and first incorporate the U
duality symmetries of M theory which lead to a relation among
$z^I_\alpha$ and to an elegant form for ${\cal G}^{I J} \;$.


\vspace{4ex}

\centerline{\bf U duality relation among $(\Pi_I, p_{\alpha I})
\;$}

\vspace{2ex}

We use the U duality symmetries here the same way as in our
earlier works \cite{k07, bdk, bk}. We note that, for a given
${\cal N}$, different intersecting brane configurations of M
theory can be related to each other by suitable U duality
operations -- namely, by suitable dimensional reduction and
uplifting to and from type IIA string theory, and T and S
dualities in type IIA/B string theories. For a metric of the
form given in equation (\ref{ds}), such an operation leads to
relations among $\lambda^\alpha$ which in turn, through their
equations of motion, imply relations among $(\Pi, p_\alpha) \;$.
Although only time dependent cases in early universe were
studied in \cite{k07, bdk, bk}, the U duality details of these
works carry over and are applicable to the static cases also
with only a few minor changes. The relevant details may be found
in \cite{k07} and in Appendix {\bf A} of \cite{bk}. Hence, we
present only the main results here without repeating the
details.

It can be shown that the relations among $(\Pi, p_\alpha)$ for
intersecting branes of M theory, obtained by applying U duality
operations as described above, are all satisfied if the
individual $(\Pi_I, p_{\alpha I})$ obey the relation
\begin{equation}\label{Ups}
p_{\parallel I} \; = \; \Pi_I + p_{0 I} + p_{\perp I} 
+ m \; (p_{\Omega I} - p_{\perp I}) 
\end{equation} 
where $p_{\parallel I}$ and $p_{\perp I}$ are the pressures
along the directions that are parallel and transverse to the
worldvolume of the $I^{th}$ stack of branes. The above relation
is a consequence of U duality symmetries and, therefore, must
always be valid independent of the details of the equations of
state. We further take $p_{\Omega I} = p_{\perp I}$ which is
natural since the sphere directions are transverse to the
branes. \footnote{ We mention here that, in the studies of early
universe or of stars, the constituent matter components also
satisfy $T^r_{\; \; r (I)} = T^a_{\; \; a (I)}$, equivalently
$\Pi_I = p_{a I} \;$. Using $p_{0 I} = - \rho_I$ and $p_{a I} =
p_{\Omega I} = p_{\perp I}$, equation (\ref{Ups}) then becomes
$p_{\parallel I} = - \rho_I + 2 p_{\perp I} \;$ \cite{cm, k07,
bdk, bk}.}

It can further be argued \cite{k07} that the equations of state
may be written in terms of one single function only. For
example, they may be written as
\begin{equation}\label{ueos}
h_{\alpha I} = \Pi_I + p_{\alpha I} = 
z_\alpha \; {\cal F} (\{*\}_I)
\; \; , \; \; \;
\alpha = \{ 0, \parallel, \perp, \Omega \} 
\end{equation}
where $\{*\}_I$ denote brane quantities such as the number and
the nett charge of the branes in $I^{th}$ stack, the constant
coefficients $z_\alpha$ and the functional dependence of ${\cal
F}$ on the brane quantities $\{*\}$ are same for all $I \;$, and
\begin{equation}\label{z}
z_\Omega = z_\perp \; \; \; , \; \; \; \; 
z_\parallel = z_0 + z_\perp 
\end{equation}
as follows from $p_{\Omega I} = p_{\perp I}$ and from equation
(\ref{Ups}). Note that equations (\ref{ueos}) may still lead to
equations of state of the type considered in equation
(\ref{eos}) if ${\cal F}$ depends on $\Pi$ only and if the
function ${\cal F}(\Pi)$ is such that, for example, ${\cal
F}(\Pi) = u^{(lo)} \Pi \; \;$ or $\; u^{(hi)} \Pi \; \;$ for low
or high magnitudes of $\Pi \;$. Equation (\ref{eos}) then
follows, with $z^I_\alpha = z_\alpha u^I$ where $u^I = u^{(lo)}$
or $u^{(hi)}$ according to whether the magnitude of $\Pi_I$ is
low or high.

Consider the case where $z^I_\alpha = z_\alpha u^I \;$ and
$z_\alpha$ obey equations (\ref{z}). It then follows from
equations (\ref{GIJ}) and from BPS intersection rules that
\begin{equation}\label{genz^alphaI}
z^{\alpha I} = \left( \frac{(n_I + 1)}{9} \; z_0 + z_\perp 
- z_\alpha \right) \; u^I 
\end{equation}
\begin{equation}\label{genGIJM}
{\cal G}^{I J} = 2 z_0 
\left( z_\perp - z_0 \; \delta^{I J} \right) \; u^I u^J \; \; .
\end{equation}
The form of $z^{\alpha I}$ and the relations among them are
consequences of U duality symmetries of M theory. The relations
among $z^{\alpha I}$ will be reflected in the solutions for
$\lambda^\alpha$ when written in terms of $l^I$, see equation
(\ref{lalphatt}) and subsequent comments. Note the elegant
structure of ${\cal G}^{I J}$ and its independence on $n_I \;$.
These features of ${\cal G}^{I J}$ are consequences of both the
U duality symmetries and the BPS intersection rules. Similar
${\cal G}^{I J}$ was also present in our time dependent early
universe studies.

\vspace{4ex}

\centerline{\bf Equations of state from action $S$}

\vspace{2ex} 

Now consider the equations of state for M theory branes which
are given in equation (\ref{rhopianswer}). They follow from the
energy momentum tensor in equation (\ref{bhtmnI}), which in turn
follows from the action $S$ given in equation (\ref{s0}).
Setting $p_{0 I} = - \rho_I$, it follows from equation
(\ref{rhopianswer}) that $h_{\alpha I} = \Pi_I + p_{\alpha I}$
are given by
\[
\left( h_{0 I}, \; h_{\parallel I}, \; h_{\perp I}, \; 
h_{\Omega I} \right) \; = \;
(2, \; 2, \; 0, \; 0) \; \Pi_I \; \; . 
\]
Clearly, these $h_{\alpha I}$ satisfy the U duality relations in
equation (\ref{Ups}). They are of the form given in equation
(\ref{ueos}) with ${\cal F} (\{*\}_I) = \Pi_I \;$ and with
$z_\alpha$ given by
\begin{equation}\label{eosM} 
(z_0, \; z_\parallel, \; z_\perp, \; z_\Omega) = 
(2, \; 2, \; 0, \; 0) 
\end{equation}
which satisfy equations (\ref{z}). Writing $h_{\alpha I} =
z^I_\alpha \; \Pi_I$ with $z^I_\alpha = z_\alpha u^I$ and $u^I =
1 \;$, it follows from equations (\ref{genz^alphaI}) and
(\ref{genGIJM}) that $z^{\alpha I}$ and ${\cal G}^{I J}$ for M
theory branes are given by
\begin{equation}\label{z^alphaI}
z^{0 I} = z^{\parallel I} = \frac{2 \; (n_I - 8)}{9} 
\; \; , \; \; \; 
z^{\perp I} = z^{\Omega I} = \frac{2 \; (n_I + 1)}{9} 
\; \; , \; \; \; 
{\cal G}^{I J} = - 8 \; \delta^{I J} \; \; .
\end{equation}
Thus $(z^\parallel, \; z^\perp) = \frac{2}{3} \; (- 2, \; 1)$
for $M2$ branes and $(z^\parallel, \; z^\perp) = \frac {2} {3}
\; (- 1, \; 2)$ for $M5$ branes.


\vspace{4ex}

\centerline{\bf 5. Solutions in terms of $\tau$}

\vspace{2ex}

We now solve the equations of motion in complete generality for
the case where the equations of state for M theory branes are
given by equations (\ref{rhopianswer}). The fact that
$z_\Omega^I = z_\perp^I = 0$ in this case leads to crucial
simplifications which in turn lead to explicit analytical
solutions.

\vspace{4ex}

\centerline{\bf Expressions for $\lambda^\alpha$ and ${\cal
A}_{(I)}$ in terms of $l^I$ and $\chi$}

\vspace{2ex}

Using $(z_0^I, z_\parallel^I, z_\perp^I, z_\Omega^I) = (2, 2, 0,
0)$ and ${\cal G}^{I J} = - 8 \; \delta^{I J}$ in equations
(\ref{lIdefn}), (\ref{lItt}), and (\ref{chitt}), we obtain
\begin{equation}\label{lIM}
l^I = 2 \; ( \lambda^0 + \sum_{i \in \parallel_I} \lambda^i )
\end{equation}
where $i \in \; \parallel_I $ denote spatial directions of the
$I^{th}$ stack of branes; and
\begin{equation}\label{lIchieqn}
l^I_{\tau \tau} = 8 \; \rho_{I 0} \; e^{l^I}
\; \; , \; \;\; 
\chi_{\tau \tau} = (m - 1)^2 \; e^{2 \chi}  
\end{equation}
where $\rho_{I 0} = - \Pi_{I 0} \;$. These simplified equations
for $l^I$ and $\chi$ can now be solved explicitly. Integrating
the above equations once, we get
\begin{equation}\label{clcchi}
(l^I_\tau)^2 - 16 \; \rho_{I 0} \; e^{l^I} = 2 \; c^I
\; \; , \; \;\; 
(\chi_\tau)^2 - (m - 1)^2 \; e^{2 \chi}  = \frac{c_\chi}{2} 
\end{equation}
where $c^I$ and $c_\chi$ are integration constants. Furthermore,
it follows from the asymptotic conditions given in equations
(\ref{bc}) and (\ref{bc2}) that, in the limit $\tau \to \tau_0
\;$,
\begin{equation}\label{bclIchi}
e^{l^I} \to 1 \; \; , \; \; \; 
e^\chi \to \frac {1} {(m - 1) \; (\tau_0 - \tau)} 
\; \; . 
\end{equation}

Using equations (\ref{lIchieqn}) for $e^{l^I}$ and $e^{2 \chi}
\;$, equation (\ref{lalphatt}) for $\lambda^\alpha$ can now be
integrated to give
\begin{equation}\label{lalpha}
\lambda^\alpha = - \; \frac{1}{8} \; \sum_I z^{\alpha I} \; l^I
+ \frac{\delta^{\alpha a}}{m - 1} \; \chi 
+ L^\alpha \; (\tau - \tau_0)
\end{equation}
where $L^\alpha$ are constants, $L^a = L^\Omega$ for all $a$,
and the asymptotic conditions given in equation (\ref{bc}) have
been incorporated. Substituting the above $\lambda^\alpha$ into
the definitions of $l^I$ and $\chi$ leads to ${\cal N} + 1$
consistency constraints on $L^\alpha \;$ given by
\begin{equation}\label{LN+1}
\sum_\alpha z^I_\alpha \; L^\alpha = 0 
\; \; , \; \; \; 
\sum_\alpha L^\alpha = L^\Omega 
\; \; . 
\end{equation}
Substituting $\lambda^\alpha$ into equation (\ref{2Lambdatau2}),
and using equations (\ref{clcchi}) and (\ref{LN+1}), leads to
one further constraint among $(c^I, c_\chi, L^\alpha)$ given by
\begin{equation}\label{consteqn}
\frac{m}{2 (m - 1)} \; c_\chi = \frac{1}{4} \; \sum_I c^I 
- \sum_{\alpha \beta} G_{\alpha \beta} \; L^\alpha \; L^\beta
\; \; . 
\end{equation}
Note that the constraint $\sum_\alpha L^\alpha = L^\Omega$
implies that
\begin{eqnarray*}
- \; \sum_{\alpha \beta} G_{\alpha \beta} L^\alpha L^\beta 
& = & - \; ( \; \sum_\alpha L^\alpha \; )^2 
+ \sum_\alpha (L^\alpha)^2 \\
& = & (L^0)^2 + \sum_i (L^i)^2 + (m - 1) \; (L^\Omega)^2
\; \; \ge 0 \; \; . 
\end{eqnarray*}
Hence $c_\chi \ge 0 \;$ if $\; \sum_I c^I \ge 0 \;$.

An expression for the field strength $F_{0 i_1 \cdots i_{n_I}
r}$ can also be obtained. Using $\rho_{I 0} = - \Pi_{I 0}$,
equation (\ref{PiIhI}) for $\Pi_I$, and equation (\ref{lIM}) for
$l^I$, it follows from equation (\ref{rhopianswer}) that
\[
\rho_I = - \Pi_I = \rho_{I 0} \; e^{- 2 \Lambda + l^I} 
= \frac{1}{4} \; e^{- 2 \lambda - l^I} \; 
(F_{0 i_1 \cdots i_{n_I} r})^2 \; \; , 
\]
and hence that $\rho_{I 0} \ge 0 \;$. Furthermore, since $F_{0
i_1 \cdots i_{n_I} r} = \partial_r {\cal A}_{(I)} \;$, it
follows that
\[
e^{2 \Lambda - 2 \lambda} \; ( \partial_r {\cal A}_{(I)} )^2 
= \left( \partial_\tau {\cal A}_{(I)} \right)^2 
= 4 \; \rho_{I 0} \; e^{2 l^I} 
\]
where the first equality follows from equation (\ref{Xtaur}).
Using the asymptotic form for ${\cal A}_{(I)}$ given in equation
(\ref{bc}), we then get
\begin{equation}\label{AIlI}
e^{\Lambda - \lambda} \; \partial_r {\cal A}_{(I)} 
= \partial_\tau {\cal A}_{(I)} 
= - 2 \; \sqrt{\rho_{I 0}} \; \;  e^{l^I} 
\end{equation}
and, taking $Q_I$ to be positive with no loss of generality,
\begin{equation}\label{rhoIQI}
2 \; \sqrt{\rho_{I 0}} = (m - 1) \; Q_I \; \; . 
\end{equation}


\vspace{4ex}

\centerline{\bf Solutions for $l^I$, $\; \chi$, and ${\cal
A}_{(I)}$ }

\vspace{2ex}

We now solve equations (\ref{lIchieqn}) for $l^I$ and $\chi
\;$. Let $X(\tau)$ satisfy the equation 
\[
X_{\tau \tau} = b \; e^X 
\; \; \Longrightarrow \; \; \;
(X_\tau)^2 - 2 b \; e^X = 2 c
\]
where $b$ is a given constant and $c$ an integration constant.
$X$ may be thought of as the coordinate of a particle moving in
a `potential' $(- b e^X)$ with `energy' $c \;$. The second
equation above implies that if $b$ is negative then $c$ must be
positive. Thus, there are three cases depending on the signs of
$b$ and $c \;$. The corresponding general solutions may be
written as
\[
e^X = \frac{2 \; \gamma^2}{\alpha^2 \; sinh^2 y} 
\; \; \; , \; \; \; \; 
b = \alpha^2 \; , \; \; c = 2 \; \gamma^2 
\]
\[
e^X = \frac{2 \; \gamma^2}{\alpha^2 \; sin^2 y} 
\; \; \; , \; \; \; \; 
b = \alpha^2 \;  , \; \; c = - \; 2 \; \gamma^2 
\]
\[
e^X = \frac{2 \; \gamma^2}{\alpha^2 \; cosh^2 y}
\; \; \; , \; \; \; \; 
b = - \alpha^2 \;  , \; \; c = 2 \; \gamma^2 
\]
where $y = \gamma (\tau_* - \tau)$, $\; \alpha$ is obtained from
$b$, and $\gamma$ and $\tau_*$ are two integration constants.
Note that, in the above three cases, $X$ is unbounded above and
below, bounded from below, or bounded from above respectively.

In the following, we take $\alpha$ and $\gamma$ to be positive
with no loss of generality. The constant $b$ is given, hence it
determines the sign of $b$ and the value of $\alpha \;$. The
constants $\gamma$ and $\tau_*$ may be determined, for example,
as follows.  Let $(x, v)$ be the initial values of $(X, X_\tau)$
at $\tau = \tau_{in} \;$. Then $v^2 - 2 b e^x = 2 c$ determines
the sign of $c$ and the value of $\gamma \;$. The solution for
$e^X$ can now be selected from above according to the signs of
$b$ and $c \;$. This solution evaluated at $\tau = \tau_{in}$
then determines $y_{in} = \gamma (\tau_* - \tau_{in})$, and thus
$\tau_* \;$. 

Now consider equations (\ref{lIchieqn}) and (\ref{clcchi}). Let
$\rho_{I 0} > 0$ and $c^I$ be positive for all $I \;$. (We will
later comment on the case where one or more of the $c^I$ may be
negative.) Then $c_\chi > 0$ as follows from equation
(\ref{consteqn}). Hence let $c^I = 2 (\gamma^I)^2$ and $c_\chi =
2 \beta^2 \;$, and take $\gamma^I$ and $\beta$ to be positive
with no loss of generality. Then, comparing with the equations
and solutions for $X$ given above, it follows that the general
solutions for $l^I$ and $\chi$ may be written as
\begin{equation}\label{lIsoln} 
e^{l^I} \; = \; \frac{(\gamma^I)^2}{4 \; \rho_{I 0} \; \; 
sinh^2 \; y_I}
\; \; , \; \; \; 
y_I = \gamma^I \; (\tau_I + \tau_0 - \tau)
\end{equation}
\begin{equation}\label{chisoln} 
e^\chi \; = \; 
\frac{\beta} {(m - 1) \; \; sinh \; y_\chi}
\; \; , \; \; \; 
y_\chi = \beta \; (\tau_0 - \tau)
\end{equation}
where $\tau_I$ are integration constants, and we have
incorporated the asymptotic condition given in equation
(\ref{bclIchi}) for $e^\chi \;$ in the limit $\tau \to \tau_0
\;$. It follows from equation (\ref{consteqn}) that $\beta$, $\;
\gamma^I$, and $L^\alpha$ must satisfy the constraint
\begin{equation}\label{bw}
\frac{m}{m - 1}  \; \beta^2 = \frac{1}{2} \; 
\sum_I (\gamma^I)^2 - \sum_{\alpha \beta} G_{\alpha \beta} \;
L^\alpha \; L^\beta \; \; .
\end{equation} 

We now express $e^{l^I}$ in a different form that facilitates
comparisons with the brane solutions as presented commonly in
the literature. Since $e^{l^I} \to 1$ in the limit $\tau \to
\tau_0 \;$, we have
\[ 
2 \; \frac{\sqrt{\rho_{I 0}}}{\gamma^I} \; \; 
sinh \; (\gamma^I \tau_I) = 1 \; \; , 
\]
which may be satisfied identically upon setting
\[
\frac{\sqrt{\rho_{I 0}}}{\gamma^I} e^{\gamma^I \tau_I} = 
{\cal C}_I^2 \; \; , \; \; \; 
\frac{\sqrt{\rho_{I 0}}}{\gamma^I} e^{- \gamma^I \tau_I} =
{\cal S}_I^2
\]
where ${\cal C}_I = cosh \; \Theta_I$ and ${\cal S}_I = sinh \;
\Theta_I \;$. It then follows that
\[
\frac{\sqrt{\rho_{I 0}}}{\gamma^I} = {\cal C}_I \; {\cal S}_I
\; \; , \; \; \; 
e^{\gamma^I \tau_I} = \frac{{\cal C}_I}{{\cal S}_I} \; \; .
\]
Using the above expressions in equation (\ref{lIsoln}), we now
obtain
\begin{equation}\label{lFH}
e^{l^I} \; = \; \frac {e^{2 \gamma^I (\tau - \tau_0)}} {H_I^2}
\; \; , \; \; \; 
H_I \; = \; {\cal C}_I^2 - e^{2 \gamma^I (\tau - \tau_0)} 
\; {\cal S}_I^2 \;\; .
\end{equation} 
Using the expression for $e^{l^I}$ given above, equation
(\ref{AIlI}) can now be integrated to obtain ${\cal A}_{(I)}
\;$. Noting that $H_I \to 1$ and ${\cal A}_{(I)} \to 0$ in the
limit $\tau \to \tau_0 \;$, and using $\frac {\sqrt{\rho_{I 0}}}
{\gamma^I} = {\cal C}_I \; {\cal S}_I \;$, it follows from
equations (\ref{AIlI}) and (\ref{lFH}) that
\begin{equation}\label{AIFH}
{\cal A}_{(I)} \; = \; \frac {{\cal C}_I \; {\cal S}_I} {H_I} \;
\left( 1 - e^{2 \gamma^I (\tau - \tau_0)} \right) \; \; ;
\end{equation}
and from equation (\ref{rhoIQI}) for the charges $Q_I$ that
\begin{equation}\label{QIgammaI}
(m - 1) \; Q_I = 2 \gamma^I \; {\cal C}_I \; {\cal S}_I \; \; .
\end{equation}
Also, using $l^I = 2 \; \gamma^I (\tau - \tau_0) - 2 \; ln H_I
\;$, the solution for $\lambda^\alpha$ given in equation
(\ref{lalpha}) can be written in terms of $H_I \;$ as
\begin{equation}\label{lalphaHI}
\lambda^\alpha \; = \; 
\sum_I \frac{z^{\alpha I}}{4} \; ln \; H_I
\; + \; \frac{\delta^{\alpha a}}{m - 1} \; \chi
\; + \; 
\left( L^\alpha - \sum_I \frac{z^{\alpha I} \; \gamma^I}{4}
\right) (\tau - \tau_0) \; \; .
\end{equation}

We note here that the functions $H_I$ and ${\cal A}_{(I)}$ and
the form of $\lambda^\alpha$ given above turn out to be the
generalisations of the corresponding harmonic functions, the
$(n_I + 1)-$form gauge fields, and the metric components which
appear in the standard intersecting brane solutions as given,
for example, in \cite{t96,alg, hm, rt97, aeh}.


\vspace{4ex}

\centerline{\bf 6. Solutions in terms of $r \;$: obtaining
$r(\tau)$}

\vspace{2ex}

We now have the complete general solutions for static
intersecting M theory branes, obtained by directly solving the
equations of motion in terms of the variable $\tau \;$. To
express these solutions in terms of the more familiar radial
coordinate $r$, we need to obtain the function $r(\tau) \;$ in
complete generality using equation (\ref{rtau}). We place an
emphasis here on the generality of $r(\tau)$ because of the
following. General solutions in terms of $\tau$ have been
obtained by several groups for systems which, at different
levels, are more general than ours \cite{ar}. But, while
translating such solutions to the $r$ variable, invariably some
ansatz is made which thereby limits the generality of the final
solutions given in terms of $r \;$.

Consider the line element $d s$ given in equation (\ref{ds}).
With no loss of generality, we exchange the function $e^\lambda$
for another function $f$ by taking
\[
e^{2 \lambda} d r^2 + e^{2 \sigma} d \Omega_m^2 \; = \; 
e^{2 \sigma} \; \left( \frac{d r^2}{r^2 f} 
+ d \Omega_m^2 \right) \; \; , 
\]
thus $e^{\sigma - \lambda} = r \sqrt{f} \;$. Using $e^\lambda \;
d r = e^\Lambda \; d \tau$ and $e^{\Lambda - \sigma} = e^\chi$,
it now follows that
\begin{equation}\label{rtauf}
r_\tau = \frac{d r}{d \tau} = 
e^{\Lambda - \lambda} = r \sqrt{f} \; e^\chi \; \; . 
\end{equation}
Since $e^\chi \to r^{m - 1}$ in the limit $\tau \to \tau_0 \;$,
see equation (\ref{bc2}), the general relation between $r$ and
$\tau$ can be expressed as
\begin{equation}\label{chih}
e^{\chi(\tau)} = r^{m - 1} \; e^{h(\tau)} 
\end{equation}
where the function $h(\tau) \to 0$ in the limit $\tau \to \tau_0
\;$. Using equation (\ref{chisoln}) for $e^\chi$, it now follows
that $r(\tau)$ is given by
\begin{equation}\label{r(tau)}
r^{m - 1} = e^{\chi- h} = \frac{\beta}{m - 1} \; \; 
\frac{e^{- h}} {sinh \; \beta \; (\tau_0 - \tau)} \; \; .
\end{equation}
Calculating $r_\tau$ from this expression and substituting it
into equation (\ref{rtauf}) gives
\begin{equation}\label{sqrtf}
\sqrt{f} = \frac{r_\tau}{r} \; e^{- \chi} =
cosh \; \beta \; (\tau_0 - \tau) - \frac{h_\tau}{\beta} \;
sinh \; \beta \; (\tau_0 - \tau) \; \; .
\end{equation}
Thus $r(\tau)$ and $f(\tau)$ are determined once $h(\tau)$ is
known. 


\vspace{4ex}

\centerline{\bf Obtaining $h(\tau)$}

\vspace{2ex}

We now obtain $h(\tau)$ first for vacuum solutions, thereby also
illustrating our method in a simpler context; and then for brane
solutions. The function $h(\tau)$ turns out to be the same in
both of these cases.

Consider first the vacuum equations of motion in terms of $r$,
namely equations (\ref{Lambdar2}) and (\ref{alpharr}) with
$\Pi_I = p_{\alpha I} = 0 \;$. Equation (\ref{alpharr}) gives
\[
\lambda^{i'} = b^{i'} \; F 
\]
where $i' = (0, i)$, $\; \lambda^{i'} = (\lambda^0, \lambda^i)$,
$\; b^{i'} = (b^0, b^i)$ are constants, and the function $F$ is
the solution to the homogeneous equation
\begin{equation}\label{Frr}
F_{r r} + (\Lambda_r - \lambda_r) \; F_r = 0 \; \; . 
\end{equation}
This equation for $F$ implies that
\begin{equation}\label{F}
e^{\Lambda - \lambda} \; F_r = F_\tau = {\cal M}
\; \; \; \Longrightarrow \; \; \; 
F = {\cal M} \; (\tau - \tau_0) 
\end{equation}
where ${\cal M} = (m - 1) \; r^{m - 1}_0$ is an integration
constant and we have incorporated the condition that $F \to 0$
in the limit $\tau \to \tau_0$, see equations (\ref{bc}) and
(\ref{bc2}). As for $\sigma$, note that there is a freedom in
defining the $r$ coordinate. Using this freedom, we set
\[
\sigma = b^\Omega \; F + ln \; r 
\]
with no loss of generality. \footnote{ Upon substitution of
$\sigma(r)$, equations (\ref{Lambdar2}) and (\ref{alpharr}) will
now give an equation for the function $f(r) \;$. See Appendix
{\bf C}.} Thus, for vacuum solutions, we have
\begin{equation}\label{vaclambdaFr}
\lambda^\alpha \; = \; b^\alpha \; F 
+ \; \delta^{\alpha a} \; \; ln \; r 
\end{equation}
where $b^\alpha = (b^0, b^i, b^a)$, and $\lambda^a = \sigma$ and
$b^a = b^\Omega$ for all $a \;$. One can further set $b^\Omega =
0$ with no loss of generality, so that $r$ becomes the
conventional radial coordinate. However, we keep $b^\Omega$ non
vanishing since the underlying structure is then clearer. Also, 
define
\begin{equation}\label{BK}
B = \sum_\alpha b^\alpha 
\; \; , \; \; \; 
K = - \sum_{\alpha \beta} G_{\alpha \beta} \; 
b^\alpha \; b^\beta \; \; .
\end{equation}
Using $\lambda^\alpha$ given in equations (\ref{vaclambdaFr}),
we now get
\[
\chi = \Lambda - \sigma = 
(m - 1) \; ln \; r + (B - b^\Omega) \; F \; \; .
\]
Comparing with equation (\ref{chih}) then gives $h = (B -
b^\Omega) F$ for vacuum solutions. 

Consider now the brane equations of motion in terms of $r$,
namely equations (\ref{Pir}) -- (\ref{alpharr}) with $\Pi_I$ and
$p_{\alpha I}$ given by equations (\ref{rhopianswer}). Following
the comments made below equation (\ref{GIJ}) regarding
expressing $\lambda^\alpha$ in terms of $l^I$, and following the
analysis of section {\bf 5}, it can be seen that $\lambda^{i'} =
(\lambda^0, \lambda^i)$ are now given in terms of $l^I$ by
equation (\ref{lalpha}), or equivalently in terms of $H_I$ by
equation (\ref{lalphaHI}), where $z^{\alpha I}$ are given in
equation (\ref{z^alphaI}) and $(\tau - \tau_0)$ is now to be
replaced by the function $F$ satisfying the homogeneous equation
(\ref{Frr}). Thus, we write
\[
\lambda^{i'} \; = \; \sum_I \frac{z^{i' I}}{4} \; ln \; H_I \; 
+ \; b^{i'} \; F \; \; .
\]
As for $\sigma$, note that $p_{a I} = p_{\Omega I} = p_{\perp
I}$ for the branes. It therefore follows that, using the freedom
in defining the $r$ coordinate, we can set
\[
\sigma \; = \; \sum_I \frac{z^{\perp I}}{4} \; ln \; H_I \; 
+ \; b^\Omega \; F + ln \; r 
\]
with no loss of generality. Thus, for brane solutions, we have
\begin{equation}\label{lambdaFr}
\lambda^\alpha \; = \; \sum_I \frac{z^{\alpha I}}{4} \; 
ln \; H_I \; + \; b^\alpha \; F 
+ \; \delta^{\alpha a} \; \; ln \; r \; \; . 
\end{equation}
Using $\lambda^\alpha$ given above, we again get
\[
\chi = \Lambda - \sigma = 
(m - 1) \; ln \; r + (B - b^\Omega) \; F 
\]
because it follows from equation (\ref{z^alphaI}) that
\[
\sum_\alpha z^{\alpha I} - z^{\Omega I} 
= (n_I + 1) \; z^{\parallel I} 
+ (8 - n_I) \; z^{\perp I} = 0 
\]
and, hence, the coefficient of the $ln H_I$ term in $\chi$
vanishes. Comparing with equation (\ref{chih}) then gives $h =
(B - b^\Omega) F$ for brane solutions also. 

Writing $F$ in terms of $\tau$ using equation (\ref{F}), we now
have
\begin{equation}\label{h}
h(\tau) = k \; (\tau - \tau_0) \; \; , \; \; \; 
k = {\cal M} \; (B - b^\Omega) 
\end{equation}
for both vacuum and brane solutions. The functions $r(\tau)$ and
$f(\tau)$ then follow from equations (\ref{r(tau)}) and
(\ref{sqrtf}), and are given by 
\begin{equation}\label{1r(tau)vac}
r^{m - 1} \; = \; \frac{\beta}{m - 1} 
\; \; \frac{e^{k \; (\tau_0 - \tau)}} 
{sinh \; \beta \; (\tau_0 - \tau)} 
\end{equation}
and
\begin{equation}\label{1sqrtfvac}
\sqrt{f} \; = \; cosh \; \beta \; (\tau_0 - \tau)
- \frac{k}{\beta} \; sinh \; \beta \; (\tau_0 - \tau) 
\; \; .
\end{equation}

Note that, in principle, one can now obtain the functions $F$
and $f$, and thereby all other functions also, in terms of
radial coordinate $r \;$ by using equations (\ref{F}),
(\ref{1r(tau)vac}), and (\ref{1sqrtfvac}). However, it turns out
that the functions $F(r)$ and $f(r)$ can be obtained more
directly. It is straightforward to obtain their equations of
motion in terms of $r \;$. We have earlier solved these
equations analytically and in complete generality, and reported
the results in \cite{k11}. For the sake of completeness here, we
briefly describe these analytical solutions and their derivation
in Appendix {\bf C}.


\vspace{4ex}

\centerline{\bf Relations between $(b^\alpha, {\cal M})$ and
$(L^\alpha, \gamma^I)$}

\vspace{2ex}

The brane solutions given in terms of $\tau$ are parametrised by
the set of constants $(L^\alpha, \gamma^I) \;$, whereas those
given in terms of $r$ are parametrised by the set $(b^\alpha,
{\cal M}) \;$. We now obtain the relations between these two
sets.  

Consider the expression for $\lambda^\alpha$ given in equation
(\ref{lambdaFr}). Using equation (\ref{chih}) to write $ln \; r
= \frac{\chi - h}{m - 1}$ in this expression, substituting $F =
{\cal M} \; (\tau - \tau_0)$ and $h = k \; (\tau - \tau_0)$ into
it, and then comparing it with equation (\ref{lalphaHI}), we get
\begin{equation}\label{balpha}
{\cal M} \; b^\alpha - \frac{\delta^{\alpha a}}{m - 1} \; k 
\; = \; L^\alpha - \sum_I \frac{z^{\alpha I} \; 
\gamma^I}{4} \; \; .
\end{equation}
Using the expressions for $z_\alpha^I$ and $z^{\alpha I}$ given
in equations (\ref{eosM}) and (\ref{z^alphaI}), the ${\cal N} +
1$ constraints on $L^\alpha$ given in equations (\ref{LN+1}) can
be shown to imply that $k = {\cal M} \; (B - b^\Omega)$, which
is already known, and that
\begin{equation}\label{wI} 
\gamma^I = {\cal M} \; w^I 
\; \; , \; \; \;
w^I = \frac{1}{2} \; \sum_\alpha z^I_\alpha \; b^\alpha 
= b^0 + \sum_{i \in \parallel_I} b^i 
\end{equation}
where $i \in \; \parallel_I$ denotes the spatial directions of
the $I^{th}$ stack of branes. Furthermore, using the $b^\alpha$
given in equation (\ref{balpha}), it can be shown that $K$
defined in equation (\ref{BK}) is now given by
\[
{\cal M}^2 \; K \; = \;
\frac{1}{2} \; \sum_I (\gamma^I)^2 
- \sum_{\alpha \beta} G_{\alpha \beta} \; L^\alpha \; L^\beta 
- \frac{m}{m - 1} \; k^2 \; \; .
\]
Then the constraint given in equation (\ref{bw}) becomes 
\begin{equation}\label{betabb} 
\beta^2 = k^2 + \; \frac{m - 1}{m} \; {\cal M}^2 \; K \; \; . 
\end{equation}

Note that equation (\ref{wI}) gives $\gamma^I (\tau - \tau_0) =
w^I \; F \;$. Equations (\ref{lFH}) and (\ref{AIFH}) then give
$e^{l^I} = \frac {e^{2 w^I F}} {H_I^2}$ and
\begin{equation}\label{HIAI} 
H_I =  {\cal C}_I^2 - e^{2 w^I F} \; {\cal S}_I^2
\; \; , \; \; \; 
{\cal A}_{(I)} =  \frac {{\cal C}_I \; {\cal S}_I} {H_I} \;
\left( 1 - e^{2 w^I F} \right) \; \; . 
\end{equation} 
Using $\gamma^I = (m - 1) \; r_0^{m - 1} \; w^I$, equation
(\ref{QIgammaI}) for the charges $Q_I$ gives
\begin{equation}\label{QIwI}
Q_I = (2 \; w^I \; {\cal C}_I \; {\cal S}_I) \; r_0^{m - 1} 
\; \; .
\end{equation}

Note also that there is a scaling freedom, namely
\[
(F, \; {\cal M}, \; b^\alpha) \; \; \longrightarrow \; \;  
(c \; F, \; c \; {\cal M}, \; c^{- 1} \; b^\alpha) 
\]
where $c$ is a constant, under which $(b^\alpha F, \; {\cal M}
b^\alpha)$ and hence the solutions $(\lambda^\alpha, \; H_I,
\; {\cal A}_{(I)})$ remain unchanged. Using this scaling
freedom, we can set, for example, $2 \; (B - b^\Omega) = 1$ with
no loss of generality.

We now count the nett number of constants among each of the sets
$(L^\alpha, \gamma^I)$ and $(b^\alpha, {\cal M}) \;$. The
constants $(L^\alpha, \gamma^I)$ are $(n_c + 2 + {\cal N})$ in
number, and must obey $({\cal N} + 1)$ constraints given in
equations (\ref{LN+1}). Hence, there is a nett total of $(n_c +
1)$ constants among $(L^\alpha, \gamma^I) \;$. Similarly,
$(b^\alpha, {\cal M})$ are $(n_c + 2 + 1)$ in number but with no
constraints. However, one can set $b^\Omega = 0$ using the
freedom in defining the $r$ coordinate; and set, for example, $2
\; (B - b^\Omega) = 1$ using the scaling freedom. Hence, there
is a nett total of $(n_c + 1)$ constants among $(b^\alpha, {\cal
M})$ also. This equality of the nett totals is a further check
that there has been no loss of generality in our derivation of
$r(\tau) \;$.


\vspace{4ex}

\centerline{\bf 7. Properties of the solutions}

\vspace{2ex}

We have now completed the derivation of the general solutions
for static intersecting M theory branes, expressed in terms of
$\tau$ or $r \;$. Below, for the sake of the reader's
convenience, we first collect the main results and the
expressions for the general solutions together in one place; and
then discuss the properties of the solutions.


\vspace{4ex}

\centerline{\bf Main results in terms of $\tau$}

\vspace{2ex}

In terms of the variable $\tau$, the general intersecting brane
solutions are given by
\begin{equation}\label{dstau} 
d s^2 
= - e^{2 \lambda^0} d t^2 + \sum_i e^{2 \lambda^i} (d x^i)^2
+ e^{2 \sigma} \; \left( e^{2 \chi} \; d \tau^2 + d \Omega_m^2
\right)
\end{equation}
and ${\cal A}_{(I)}$ where 
\[
\lambda^\alpha = 
\sum_I \frac{z^{\alpha I}}{4} \; ln \; H_I
+ \frac{\delta^{\alpha a}}{m - 1} \; \chi
+ \tilde{L}^\alpha \; (\tau - \tau_0) \; \; ,
\]

\[
e^\chi = \frac{\beta} {(m - 1) \; 
sinh \; \beta \; (\tau_0 - \tau)} 
\; \; \; , \; \; \; \; 
\tilde{L}^\alpha = L^\alpha 
- \sum_I \frac{z^{\alpha I} \; \gamma^I}{4} \; \; ,
\]

\[
H_I = {\cal C}_I^2 - e^{2 \gamma^I (\tau - \tau_0)} 
\; {\cal S}_I^2 
\; \; \; , \; \; \; \; 
{\cal A}_{(I)} = \frac {{\cal C}_I \; {\cal S}_I} {H_I} \;
\left( 1 - e^{2 \gamma^I (\tau - \tau_0)} \right) \; \; , 
\]
and the constants $z^{\alpha I}$ are given in equation
(\ref{z^alphaI}). In the above expressions, $(\beta, \gamma^I)$
are taken to be positive, $\beta$ is given by equation
(\ref{bw}), and $L^\alpha$ obey the constraints given in
equation (\ref{LN+1}). The general $Mp$ brane solutions follow
upon setting ${\cal N} = 1 \;$. Thus, for example, the line
element for the general $Mp$ brane solutions is given by
\begin{equation}\label{mptau} 
d s^2_{(Mp)} = H^{A^\parallel} \; d s^2_{p + 1} 
+ H^{A^\perp} e^{2 \tilde{\sigma}} \; 
\left( e^{2 \chi} \; d \tau^2 + d \Omega_{9 - p}^2
\right)
\end{equation} 
where the $I$ subscripts are omitted, 
\[ 
d s^2_{p + 1} = - e^{2 \tilde{L}^0 (\tau - \tau_0)} \; d t^2 
+ \sum_{i = 1}^p e^{2 \tilde{L}^i (\tau - \tau_0)} \; 
(d x^i)^2 \; \; ,
\]
\[
e^{2 \tilde{\sigma}} \; = \; 
e^{\frac{2 \; \chi}{8 - p}} \; \; 
e^{2 \tilde{L}^\Omega \; (\tau - \tau_0)} \; 
\; \; \; , \; \; \; \; 
\tilde{L}^\alpha = L^\alpha - \frac{z^\alpha \; \gamma}{4} 
\; \; ,
\] 
and $(p, \; A^\parallel, \; A^\perp) = (2, \; \frac{- 2}{3}, \;
\frac{1}{3})$ for $M2$ branes and $(5, \; \frac{- 1}{3}, \;
\frac{2}{3})$ for $M5$ branes. The general vacuum solutions
follow upon taking ${\cal S}_I = 0 \;$, thus $H_I = 1$ and
${\cal A}_{(I)} = 0 \;$, and formally setting $\gamma^I = 0$ in
the expression for $\tilde{L}^\alpha \;$.


\vspace{4ex}

\centerline{\bf Main results in terms of $r$}

\vspace{2ex}

In terms of the radial coordinate $r$, the general intersecting
brane solutions are given by
\begin{equation}\label{dsr} 
d s^2 = - e^{2 \lambda^0} d t^2 + \sum_i e^{2 \lambda^i} 
(d x^i)^2 + e^{2 \sigma} \; \left( \frac{d r^2}{r^2 f} 
+ d \Omega_m^2 \right)
\end{equation} 
and ${\cal A}_{(I)}$ where
\[
\lambda^\alpha \; = \; \sum_I \frac{z^{\alpha I}}{4} \; 
ln \; H_I \; + \; b^\alpha \; F 
+ \; \delta^{\alpha a} \; \; ln \; r \; \; ,
\]
and 
\[
H_I =  {\cal C}_I^2 - e^{2 w^I F} \; {\cal S}_I^2
\; \; , \; \; \; 
{\cal A}_{(I)} \; = \; \frac {{\cal C}_I \; {\cal S}_I} {H_I} \;
\left( 1 - e^{2 w^I F} \right) \; \; .
\]
The functions $F$ and $f$ in these expressions are not obtained
in terms $r \;$. Instead $r$, $\; F$, and $f$ are all obtained
in terms of $\tau$ and are given by
\begin{equation}\label{r(tau)vac}
r^{m - 1} = \frac{\beta \; e^{k \; (\tau_0 - \tau)}} 
{(m - 1) \; sinh \; \beta \; (\tau_0 - \tau)}
\; \; \; , \; \; \; \; 
F = {\cal M} \; (\tau - \tau_0) \; \; , 
\end{equation}

\begin{equation}\label{sqrtfvac}
\sqrt{f} = cosh \; \beta \; (\tau_0 - \tau) - \frac{k}{\beta} 
\; sinh \; \beta \; (\tau_0 - \tau) \; \; ,
\end{equation}
using which one can obtain $F(r)$ and $f(r)$, see the comments
below equation (\ref{1sqrtfvac}). In the above expressions,
$(\beta, w^I, k)$ are taken to be positive,
\[
{\cal M} = (m - 1) \; r^{m - 1}_0
\; \; \; , \; \; \; \; 
\beta^2 = k^2 + \; \frac{m - 1}{m} \; {\cal M}^2 \; K \; \; ,
\]
and $(w^I, k, K)$ are defined in equations (\ref{wI}),
(\ref{h}), and (\ref{BK}). The general $Mp$ brane solutions
follow upon setting ${\cal N} = 1 \;$. Thus, for example, the
line element for the general $Mp$ brane solutions is given by
\begin{equation}\label{mpr} 
d s^2_{(Mp)} = H^{A^\parallel} \; d s^2_{p + 1} 
+ H^{A^\perp} \; e^{2 b^\Omega F} \; \left( \frac{d r^2}{f}
+ r^2 d \Omega_{9 - p}^2 \right) 
\end{equation} 
where the $I$ subscripts are omitted, 
\[
d s^2_{p + 1} = - e^{2 b^0 F} d t^2 
+ \sum_{i = 1}^p e^{2 b^i F} (d x^i)^2 \; \; , 
\]
and $(p, \; A^\parallel, \; A^\perp) = (2, \; \frac{- 2}{3}, \;
\frac{1}{3})$ for $M2$ branes and $(5, \; \frac{- 1}{3}, \;
\frac{2}{3})$ for $M5$ branes. The general vacuum solutions
follow upon taking ${\cal S}_I = 0 \;$, thus $H_I = 1$ and
${\cal A}_{(I)} = 0 \;$.


\vspace{4ex}

\centerline{\bf Behaviour of $(F, \; f, \; r)$ and 
$(H_I, \; {\cal A}_{(I)})$}

\vspace{2ex}

We now describe the behaviour of the functions $(F, \; f, \; r)$
and $(H_I, \; {\cal A}_{(I)})$ as $\tau$ decreases from $\tau_0$
to $- \infty \;$. The behaviour of $\lambda^\alpha$ then
follows. It is clear from their expressions that these functions
remain finite and vary smoothly in the open range $- \infty <
\tau < \tau_0 \;$. Therefore, we will focus on their behaviours
in the limits where $\tau \to \tau_0$ and where $\tau \to -
\infty \;$. In the following, we set $b^\Omega = 0$ and $2 \; (B
- b^\Omega) = 1$ with no loss of generality. Then
\[ 
B = b^0 + \sum_i b^i = \frac{1}{2} 
\; \; , \; \; \; 
2 \; k = {\cal M} \; \; , \; \; \; 
K = (b^0)^2 + \sum_i (b^i)^2 - \frac{1}{4} \; \; . 
\] 

\vspace{4ex}

\centerline{\bf Asymptotic limit $\tau \to \tau_0$}

\vspace{2ex}

By construction, the limit $\tau \to \tau_0$ corresponds to
asymptotically flat spacetime. In this limit, it follows from
equation (\ref{bc2}) that
\[
r^{m - 1} \; \to \; \frac {1} {(m - 1) (\tau_0 - \tau)} 
\; \to \; \infty \; \; . 
\]
Hence ${\cal M} \; (\tau_0 - \tau) \to \frac {r_0^{m - 1}} {r^{m
- 1}} \;$. It then follows that
\[
\sqrt{f} \; \to \; 1 - k \; (\tau_0 - \tau)
\; \; \;  \Longrightarrow \; \; \; 
f \; \to \;  1 - \frac{r_0^{m - 1}} {r^{m - 1}} 
\]

\[
e^F \; \to \; 1 + {\cal M} \; (\tau - \tau_0) 
\; \to \; 1 - \frac{r_0^{m - 1}} {r^{m - 1}} 
\]

\[
H_I \; \to \; 1 + (2 w^I {\cal S}_I^2) \; 
\frac{r_0^{m - 1}} {r^{m - 1}} 
\; \; \; \; \; , \; \; \; \; \; \; \; \; 
{\cal A}_{(I)} \; \to \; \frac{Q_I} {r^{m - 1}} 
\]
where $Q_I = (2 w^I {\cal C}_I {\cal S}_I) \; r^{m - 1}_0$, see
equation (\ref{QIwI}). Thus, the general brane solutions here
all behave the same way as the standard ones in the asymptotic
limit where $r \to \infty \;$. 


\vspace{4ex}

\centerline{\bf Functions $e^F$, $\; H_I$, and ${\cal A}_{(I)}$}

\vspace{2ex}

We now describe the behaviour of the functions $e^F$, $\; H_I$,
and ${\cal A}_{(I)}$ as $\tau$ decreases from $\tau_0$ and
approaches the limit $\tau \to - \infty \;$.

The behaviour of the function $e^F$ is straightforward to
see. Since $F = {\cal M} (\tau - \tau_0)$, it follows that $e^F$
varies smoothly and monotonically as $\tau$ decreases, and that
$e^F \to 0$ in the limit $\tau \to - \infty \;$. The function
$e^{2 w^I F}$ behaves the same way as $e^F \;$ since $w^I$ is
positive. Therefore, it follows that the functions $H_I$ and
${\cal A}_{(I)}$ vary smoothly and monotonically as $\tau$
decreases and, further, that $H_I \to {\cal C}_I^2$ and ${\cal
A}_{(I)} \to \frac{{\cal S_I}} {{\cal C_I}} \;$ in the limit
$\tau \to - \infty \;$.

\vspace{4ex}

\centerline{\bf Functions $r(\tau)$ and $f(\tau)$}

\vspace{2ex}

We now describe the behaviour of the functions $r(\tau)$ and
$f(\tau)$ as $\tau$ decreases away from $\tau_0$ and approaches
the limit $\tau \to - \infty \;$. Their behaviour depends on
whether $K = 0$ or $K > 0$ or $K < 0 \;$. We describe these
cases seperately.

\vspace{2ex}

\centerline{\bf $\mathbf K = 0$} 

\vspace{2ex}

Note that $\beta = k$ when $K = 0 \;$. Then, equations
(\ref{r(tau)vac}) and (\ref{sqrtfvac}) give
\[
r^{m - 1} = \frac{2 k} 
{(m - 1) \; (1 - e^{- 2 k \; (\tau_0 - \tau)})}
\; \; \; , \; \; \; \; 
\sqrt{f} = e^{- k \; (\tau_0 - \tau)} \; \; . 
\]
Hence, $\; f \to 0$ and $r \to r_0$ as $\tau \to - \infty \;$.
Indeed, using $2 k = {\cal M} = (m - 1) \; r_0^{m - 1}$, it
follows that
\[
f \; = \; e^{2 k \; (\tau - \tau_0)} \; = \; e^F \; = \; 
1 - \frac{r_0^{m - 1}} {r^{m - 1}} \; \; .  
\]
The standard intersecting brane solutions as given, for example,
in \cite{t96, alg, hm, rt97, aeh} all follow from this $K = 0$
case upon further choosing $2 b^0 = 1$ and $b^i = 0 \;$. One
then has $2 w^I = 1$ for all $I$, see equation (\ref{wI}), and
hence $e^{2 w^I F} = e^F$ and $Q_I = ({\cal C}_I {\cal S}_I) \;
r^{m - 1}_0 \;$. It then follows that $H_I$ and ${\cal A}_{(I)}$
are given by
\[
H_I \; = \; 1 + 
\frac{r_0^{m - 1} \; {\cal S}_I^2 \; } {r^{m - 1}} 
\; \; \; \; \; , \; \; \; \; \; \; \; \; 
{\cal A}_{(I)} \; = \; \frac {Q_I}{H_I \; r^{m - 1}} \; \; . 
\]

More general static intersecting brane solutions have been
obtained earlier by several groups in different forms and using
different methods \cite{zz}. They can all be shown \cite{k05} to
follow for other choices of $b^\alpha$, but still with $K = 0
\;$. The solutions in \cite{k05} were obtained by taking $f =
e^F = 1 - \frac {r_0^{m - 1}} {r^{m - 1}}$ as an ansatz; $K = 0$
was then found as a constraint. The present solutions are
completely general. They are obtained without making any ansatz
and, hence, are applicable when $K \ne 0$ also.

\vspace{2ex}

\centerline{\bf $\mathbf K > 0$} 

\vspace{2ex}

Note that $\beta > k$ when $K > 0 \;$. Then, equation
(\ref{r(tau)vac}) implies that, in the limit $\tau \to -
\infty$, 
\[ 
r^{m - 1} \; \to \; \frac{2 \; \beta}{m - 1} \; 
e^{(k - \beta) \; (\tau_0 - \tau)} \; \to \; 0 \; \; .  
\]
Consider equation (\ref{sqrtfvac}). It implies that, as $\tau$
decreases from $\tau_0$ to $- \infty$, $\; \sqrt{f}$ begins to
decrease from $1$, but remains strictly positive since $cosh (*)
> \frac {k} {\beta} \; sinh (*)$ when $\beta > k \;$. It also
implies that, in the limit $\tau \to - \infty$, 
\[ 
\sqrt{f} \; \to \; \left( 1 - \frac{k}{\beta} \right) \;
\frac{e^{\beta (\tau_0 - \tau)}}{2} \; \to \; \infty \; \; .  
\]
It therefore follows that $f(\tau)$ decreases from $1$; remains
strictly positive; reaches a non zero, positive minimum; and
then increases to $\infty$ as $\tau$ decreases from $\tau_0$ to
$- \infty \;$, equivalently as $r(\tau)$ decreases from $\infty$
to $0 \;$. It can be shown easily that $f$ reaches its minimum
$f_{min} = 1 - \frac {k^2} {\beta^2}$ at $\tau_{fmin}$ where
$\tau_{fmin}$ is given by $tanh \; \beta \; (\tau_0 -
\tau_{fmin}) = \frac {k} {\beta} \;$.

Thus, for $K > 0$, we have that $f$ remains strictly positive,
$H_I$ remain $> 1$ and finite, and $0 \le e^F \le \infty$ in the
interval $0 \le r \le \infty \;$. Therefore, it follows that the
$K > 0$ solutions have no horizons. These general, horizonless,
$K > 0$ solutions are new and, to the best of our knowledge,
have not appeared in the literature. These are the ones referred
to as `a class of new static solutions' in the title of this
paper.

\vspace{2ex}

\centerline{\bf $\mathbf K < 0$} 

\vspace{2ex}

Note that $\beta < k$ when $K < 0 \;$. Then, equation
(\ref{r(tau)vac}) implies that, in the limit $\tau \to -
\infty$,
\[
r^{m - 1} \; \to \; \frac{2 \; \beta}{m - 1}  \; 
e^{(k - \beta) \; (\tau_0 - \tau)} \; \to \; \infty \; \; .  
\]
Since $r^{m - 1} \to \infty$ in the limit $\tau \to \tau_0$, it
now follows that $r(\tau)$ decreases from $\infty$; reaches a
non zero, positive minimum; and then increases to $\infty$ again
as $\tau$ decreases from $\tau_0$ to $- \infty \;$. It can be
seen from equation (\ref{rtauf}), or by a direct calculation,
that $r$ reaches its minimum $r_{min}$ at $\tau_{rmin}$ where
$f(\tau_{rmin}) = 0 \;$.

Consider equation (\ref{sqrtfvac}). It implies that, as $\tau$
decreases from $\tau_0$ towards $- \infty$, $\; \sqrt{f}$
decreases and must reach zero at a finite value $\tau_{rmin}$
given by $tanh \; \beta \; (\tau_0 - \tau_{rmin}) = \frac
{\beta} {k} \;$. The right hand side of equation
(\ref{sqrtfvac}) then becomes negative for $\tau < \tau_{rmin}
\;$. However, we do not fully understand the solution for $\tau
< \tau_{rmin} \;$. We mention this $K < 0$ case here only for
the sake of completeness. Since it is not clear to us how to
continue the solution for $\tau < \tau_{rmin}$, nor how to
analyse the equations of motion for the corresponding parameter
values, we will not discuss the $K < 0$ case any further in this
paper.


\vspace{4ex}

\centerline{\bf 8. Remarks on the solutions}

\vspace{2ex}

We will now make several remarks, mostly in the context of
vacuum solutions. Since the functions $H_I$ and ${\cal A}_{(I)}$
are finite for $\tau \le \tau_0$, these remarks may easily be
adapted for the general intersecting brane solutions also.

Consider general static vacuum solutions. The corresponding line
element is given by
\begin{equation}\label{dsrvac} 
d s^2_{(Mp)} = - e^{2 b^0 F} d t^2 
+ \sum_{i = 1}^{n_c} e^{2 b^i F} (d x^i)^2 + e^{2 b^\Omega F} \;
\left( \frac{d r^2}{f} + r^2 d \Omega_m^2 \right)
\end{equation} 
where $r$, $\; F$, and $f$ are given by equations
(\ref{r(tau)vac}) and (\ref{sqrtfvac}).

{\bf (1) } 
Note that $f = e^F = 1 - \frac{r_0^{m - 1}} {r^{m - 1}}$ for $K
= 0 \;$, and that the standard black $n_c-$brane solution is
obtained upon further choosing $2 b^0 = 1$ and $b^i = b^\Omega =
0 \;$. It is well known \cite{rt97} that one can start with this
black brane solution, boost it along a compact direction to
first generate a momentum charge, and then use the boosted black
brane to generate various ${\cal N} = 1$ brane solutions by
suitable U duality operations -- namely, by suitable dimensional
reduction and uplifting, and T and S dualities. Further repeats
of boosting and U duality operations will generate other ${\cal
N} > 1$ intersecting brane solutions. The standard static
intersecting brane solutions of string theory can also be
generated in this way.

More general brane solutions have been obtained earlier by
several groups in different forms and using different methods
\cite{zz}. We have shown in \cite{k05} that these solutions can
all be obtained by suitable boosting and U duality operations,
where now the starting vacuum solution is as given in equation
(\ref{dsrvac}), and still with $K = 0$, but with otherwise
arbitrary values of $b^\alpha \;$.

It turns out that the general static intersecting brane
solutions presented here can also be obtained by suitable
boosting and U duality operations. The starting vacuum solution
is as given in equation (\ref{dsrvac}), but with arbitrary
values of $b^\alpha$ irrespective of whether $K = 0$ or not. We
have presented in \cite{k11} the solutions generated by this
method, and it is easy to verify that they are same as the ones
given in this paper which have been obtained by directly solving
the equations of motion in complete generality. Also, as shown
in \cite{k11}, the expressions for $(B - b^\Omega)$ and $K$
remain invariant under U duality operations. Hence, their values
remain unchanged for all the solutions related by U duality
operations.

\vspace{2ex}

{\bf (2) } 
The appearance of hyperbolic functions in the solutions for
$e^{l^I}$, and thus for $H_I$, can now be related to the
underlying boost operations mentioned above. In section {\bf 5}
here, the $Sinh$ functions arise because the integration
constants $c^I$ in equation (\ref{clcchi}) are taken to be
positive. If $c^I$ were negative then trigonometric functions
would have appeared in the solutions.

Such solutions involving trigonometric functions can also be
generated by U duality methods. Instead of a boost along a
compact $x^i$ direction, one now performs a rotation in the
$(x^i, x^j)$ directions, and proceeds with other U duality
operations which will generate `tilted brane' solutions, see
\cite{angle} for example. The underlying rotation will lead to
the appearance of trigonometric functions in the solutions. In
the present context, they correspond to choosing one or more of
the integration constants $c^I$ to be negative.

\vspace{2ex}

{\bf (3) } 
Upon compactifying on the $n_c$ dimensional torus described by
the coordinates $x^i$, one obtains an effective $d = m + 2$
dimensional non compact spacetime described by the coordinates
$x^\mu = (t, r, \theta^a) \;$. The corresponding effective
action and the line element may be written, symbolically and in
the standard notation, as
\begin{eqnarray*}
S_{(d)} & \sim & \int d^d x \; \sqrt{- g_{(d)}} \;
e^{\Lambda^c} \; \left( {\cal R}_{(d)} 
+ \sum_i (\partial_\mu \lambda^i)^2 + \cdots \right) 
\; \; \; , \; \; \; \; 
\Lambda^c = \sum_i \lambda^i 
\\ 
& \sim & \int d^d x \; \sqrt{- \hat{g}_{(d)}} \;
\left( \hat{{\cal R}}_{(d)} 
+ \sum_i (\partial_\mu \lambda^i)^2 + \cdots \right) 
\end{eqnarray*} 
and 
\[
d s^2_{(d)} = g_{(d) \; \mu \nu} \; d x^\mu \; d x^\nu =
- e^{2 \lambda^0} d t^2 + e^{2 \lambda} d r^2 
+ e^{2 \sigma} d \Omega_m^2
\]
where $g_{(d) \; \mu \nu}$ and ${\cal R}_{(d)}$ are the $d$
dimensional metric and the corresponding Ricci scalar in the
`physical frame', and $\hat{g}_{(d) \; \mu \nu} = e^{\frac {2 
\Lambda^c} {d - 2} } \; g_{(d) \; \mu \nu}$ and $\hat{{\cal
R}}_{(d)}$ are those in the `Einstein frame'. The $d$
dimensional action $S_{(d)}$ now contains $n_c$ number of
scalars descending from the sizes of the compact dimensions, and
${\cal N}$ number of $1-$form gauge fields descending from the
$(n_I + 1)-$form gauge fields ${\cal A}_{(I)} \;$. The couplings
of the scalars among themselves and with the gauge fields are
not the most general possible ones, but are dictated by the
eleven dimensional parent action $S$ given in equation
(\ref{s0}).

Clearly, the solutions presented in this paper lead to the most
general static spherically symmetric solutions of the above $d$
dimensional system with scalars and $1-$form gauge fields. For
example, Janis--Newman--Winicour--Wyman (JNWW) solutions
\cite{jnw} follow now as a special case. These JNWW solutions
can be viewed as Einstein frame solutions for the four
dimensional Brans--Dicke theory or, equivalently, as solutions
for a five dimensional theory with one compact coordinate; and,
can be shown to follow from the present vacuum solutions upon
setting $m = 2$, $\; n_c = 1$, $\; b^\Omega \ne 0$ and $K = 0$,
see \cite{k05} also.

\vspace{2ex}

{\bf (4) } 
For $K > 0$, we have that $f$ remains strictly positive, $H_I$
remain $> 1$ and finite, and $0 \le e^F \le \infty$ in the
interval $0 \le r \le \infty \;$. Thus, the $K > 0$ solutions
have no horizons. We note here that similar horizonless
solutions have also appeared recently in several works
\cite{sbw}. The set ups and the contexts of these works are
totally different from the present ones, and the reasons for the
similarities of the solutions are not clear to us.

\vspace{2ex}

{\bf (5) } 
The ADM mass $M_{ADM}$ for the solutions can be obtained from
the expressions given in Appendix {\bf B}. It is easy to see
that $M_{ADM}$ is positive for all the general intersecting
brane solutions if $b^\alpha$ satisfy the inequality 
\[
2 \; (m b^\Omega + \sum_i b^i) < \frac {m} {m - 1} \; \; . 
\]
Using $2 (B - b^\Omega) = 1$, the above inequality becomes $
b^\Omega < b^0 + \frac {1} {2 \; (m - 1)} \; $. We will assume
that $b^\alpha$ are choosen, {\em e.g.} $b^\Omega = 0$, $\; 2 B
= 1$, and $b^0 > 0 \;$, such that ADM mass is positive.

Consider now the curvature tensors. Note that the Riemann tensor
components $\hat{R}_{A B C D}$, calculated in the local tangent
frame with indices $A, B, \cdots$, correspond to tidal forces.
The non vanishing components of $\hat{R}_{A B C D}$ for the
metric in equation (\ref{ds}) are given in Appendix {\bf A}. If
any of these components diverges then we assume conservatively
that there will generically be a curvature singularity. Now, by
construction, the limit $\tau \to \tau_0$ corresponds to
asymptotically flat spacetime. Therefore, the components of
$\hat{R}_{A B C D}$ will all vanish in this limit. Furthermore,
the fields behave smoothly and remain finite for all finite
values of $\tau < \tau_0 \;$; hence, so will the $\hat{R}_{A B C
D}$ components for these values of $\tau \;$.

Consider the limit $\tau \to - \infty \;$. Let the radial
coordinate $ r \to r_S$ in this limit, hence $r_S = r_0$ if $K =
0$ and $r_S = 0$ if $K > 0 \;$. Using the expressions given in
Appendix {\bf A}, it can be seen after some algebra that, for
the general static intersecting brane solutions presented here,
except for those with $2 b^0 = 1$ and $b^i = b^\Omega = 0 \;$,
atleast one of the components of $\hat{R}_{A B C D}$ diverges in
the limit $r \to r_S \;$. Therefore, for generic values of
$b^\alpha$, there is likely to be a curvature singularity at
$r_S \;$. Note that this singularity is not covered by any
horizon, is naked, and is present even though the ADM mass is
positive. This feature, that naked singularity is present even
though ADM mass is positive, is also exhibited, for example, by
JNWW solutions, by those in \cite{zz, k05}, and by the recent
ones found in \cite{sbw}.

\vspace{2ex}

{\bf (6) } 
We will list a few reasons as to why the general static
solutions given in this paper are useful and are also likely to
be physically relevant despite the singularities at $r_S$
mentioned above. The present solutions are obtained by solving
the equations of motion in complete generality. Also, they have
positive ADM mass. Therefore, it is physically reasonable to
expect that these solutions describe the general static
configuration of the fields outside a static star in a $D = n_c
+ m + 2$ dimensional theory with compact $n_c$ dimensional
toroidal space or, equivalently, in a $d = m + 2$ dimensional
theory with $n_c$ number of scalars.

Some of the reasons for the usefulness and the physical
relevance of the present solutions are as follows.

{\bf (i) } 
In the studies of the static or dynamical properties of a star,
one must assume a most general initial field configuration in
the exterior of the star, which corresponds to the most general
solutions to the equations of motion; otherwise, one is likely
to miss some crucial properties. Any restriction of the
parameter ranges must then come from physical criteria, such as
stability of the resulting static configuration of a star or as
an attractor basin during its dynamical evolution; or from
requiring consistency with observations.

{\bf (ii) } 
Let $r_*$ be the equilibrium radius of a static star. If $r_* >
r_S$, which is trivially valid for $K > 0$ case since $r_S = 0$,
then the singularities at $r_S$ mentioned above are not relevant
since the equations of motion are different inside the star,
namely for $r < r_* \;$, and the present solutions are not
applicable there. For $r \le r_*$, one then has to solve a
different set of equations of motion which includes the stellar
material, and then match the interior solutions onto the most
general exterior ones. Further stability studies of the
resulting static configuration of the star may put an upper
bound on its mass, but they may or may not further restrict the
allowed values of $b^\alpha \;$.

{\bf (iii) } 
If a star is sufficiently massive then it is likely to
collapse. Let us assume that a static configuration is reached
at the end of such a collapse, and that the $r = r_S$ surface is
exposed. Then the present solutions are applicable for $r \ge
r_S$ and, generically, a naked singularity is likely to be
present at $r_S \;$.

It is now possible that the collpase dynamics is such that,
irrespective of the initial values of $b^\alpha$ prior to the
collapse, the final values after the collapse are always given
by an `attractor basin' where $2 b^0 = 1$ and $b^i = b^\Omega =
0 \;$ so that the $r = r_S$ surface is a standard non singular
horizon. This may be accomplished, for example, if all the non
trivial field configurations are `radiated away' during the
collapse. A singularity, similar to the one mentioned above or a
different dynamical one, may or may not occur and it may or not
may become naked during the process of collapse itself.

In the context of four dimensional JNWW solutions, such collapse
processes and various issues involving naked singularities have
been extensively studied in \cite{narayan, gj}. Similar studies
are needed in the present context also.

{\bf (iv) }
Note that the present solutions are obtained from a low energy
effective action involving only two derivative terms. Such an
action is unlikely to be applicable in the strong curvature
regime; then, one has to invoke the underlying fundamental
theory, namely string/M theory here. It is then reasonable to
assume that singularities will be resolved within such a
theory. If this is the case then the present solutions are
indeed physically relevant, but are applicable only until the
strong curvature regime is reached.

{\bf (v) } 
On the other hand, and more practically, one may simply assume
that our universe is $(n_c + 4)$ dimensional with compact $n_c$
dimensional toroidal space, or that it is four dimensional with
$n_c$ number of scalars; and that the present vacuum solutions
describe the general static configuration of the fields outside
a static star. It is then possible to study the motions of
various probes around such stars, their orbitals, et cetera and
compare them with observations. Such studies may lead to novel
phenomena or other observational consequences. See \cite{panda}
for such studies in the context of higher dimensions and $Dp$
branes, and \cite{pankaj, narayan} in the context of four
dimensional JNWW solutions.


\vspace{4ex}

\centerline{\bf 9. Conclusions}

\vspace{2ex}

We now give a brief summary and conclude by mentioning a few
issues for further studies. In this paper, we obtained
analytically the most general static intersecting brane
solutions. These solutions reduce to the known ones in special
cases, and contain further a class of new static solutions. We
described the properties of these general solutions and
discussed their usefulness and physical relevance. Along the
way, we also described the features of the brane energy momentum
tensors, the equations of motion, and their solutions which
arise as consequences of the BPS intersection rules and the U
duality symmetries of M theory.

We now mention a few issues which can be studied further. 

Assuming that the present general solutions describe the fields
outside the stars in our universe, one may calculate the motions
of probes around such stars and study the observational
consequences, as in \cite{will, pankaj, narayan, panda} for
example.

As mentioned in the Introduction, it is natural to expect that
the black holes described in M theory have been formed by the
collapse of sufficiently massive stars. It is desireable to
study such collapse processes. Towards this goal, one may first
construct a static star in M theory where the present general
solutions describe its exterior fields.

Once the static solutions for stars in M theory are obtained,
their stability properties can then be studied along the lines
of \cite{shapiro}. It is likely that a sufficiently massive star
will collapse. Such a collapse may also be studied. In these
studies of collapse, the M theoretic cosmological solutions we
had obtained earlier in \cite{k07, bdk, bk} are likely to be
useful


\vspace{4ex}

\centerline{\bf Appendix A : Riemann Tensor components}

\vspace{2ex}

Consider the metric given in equation (\ref{ds}) where $i = 1,
2, \cdots, n_c$ and the fields $(\lambda^0, \lambda^i, \lambda,
\sigma)$ depend on $r$ only. Let
\[
d \Omega_m^2 = h_{a b} \; d \theta^a \; d \theta^b
\]
where $a = 1, 2, \cdots, m$ and $h_{a b}$ depend on $\theta$s
only. Let  $e^M_{\; \; A}$ denote the $D-$bein fields and let
\[
\hat{R}_{A B C D} =
e^M_{\; \; A} \; e^N_{\; \; B} \; e^P_{\; \; C} \; e^Q_{\; \; D}
\; \; R_{M N P Q} 
\]
where $R_{M N P Q}$ is the Riemann tensor. 
It follows after a straighforward algebra that the non vanishing
components of $\hat{R}_{A B C D}$ for the metric in equation
(\ref{ds}) are given by
\begin{eqnarray*}
\hat{R}_{ r i' r j'} & = & - \delta_{i' j'} \; \;
e^{- 2 \lambda} \left( \lambda_{r r}^{i'} 
+ (\lambda_r^{i'} - \lambda_r) \; \lambda_r^{i'} \right)
\\ 
\hat{R}_{ r a r b} & = & - h_{a b} \; \; e^{- 2 \lambda} \left(
\sigma_{r r} + (\sigma_r - \lambda_r) \; \sigma_r \right)
\\ 
\hat{R}_{i' j' k' l'} & = & (\delta_{i' l'} \delta_{j' k'} -
\delta_{i' k'} \delta_{j' l'}) \; \;
e^{- 2 \lambda} \left( \lambda_r^{i'} \lambda_r^{j'} \right)
\\ 
\hat{R}_{i' a j' b} & = & - \delta_{i' j'} h_{a b} \; \;
e^{- 2 \lambda} \lambda_r^{i'} \sigma_r \\ 
\hat{R}_{a b c d} & = & e^{- 2 \sigma} \rho_{a b c d}(h) 
+ (h_{a d} h_{b c} - h_{a c} h_{b d}) \; \; 
e^{- 2 \lambda} \sigma_r^2  
\end{eqnarray*}
where $i' = (0, i)$, $\; \lambda^{i'} = (\lambda^0, \lambda^i)$,
and $\rho_{a b c d}(h)$ is the Riemann tensor corresponding to
the metric $h_{a b} \;$. 

The Riemann tensor components in terms of the variable $\tau$
defined in equation (\ref{rtau}) may be obtained from the above
expressions by replacing $(r, \lambda)$ with $(\tau, \Lambda)
\;$. Thus,
\begin{eqnarray*}
\hat{R}_{ \tau i' \tau j'} & = & - \delta_{i' j'} \; \; 
e^{- 2 \Lambda} \left( \lambda_{\tau \tau}^{i'} 
+ (\lambda_\tau^{i'} - \Lambda_\tau) \; \lambda_\tau^{i'}
\right) \\ 
\hat{R}_{ \tau a \tau b} & = & - h_{a b} \; \; e^{- 2 \Lambda}
\left( \sigma_{\tau \tau} + (\sigma_\tau - \Lambda_\tau) \;
\sigma_\tau \right) \\ 
\hat{R}_{i' j' k' l'} & = & (\delta_{i' l'} \delta_{j' k'} 
- \delta_{i' k'} \delta_{j' l'}) \; \; e^{- 2 \Lambda} \left(
\lambda_\tau^{i'} \lambda_\tau^{j'} \right) \\ 
\hat{R}_{i' a j' b} & = & - \delta_{i' j'} h_{a b} \; \;
e^{- 2 \Lambda} \lambda_\tau^{i'} \sigma_\tau 
\\ 
\hat{R}_{a b c d} & = & e^{- 2 \sigma} \rho_{a b c d}(h) 
+ (h_{a d} h_{b c} - h_{a c} h_{b d}) \; \; 
e^{- 2 \Lambda} \sigma_\tau^2  \; \; . 
\end{eqnarray*}


\vspace{4ex}

\centerline{\bf Appendix B : ADM Mass}

\vspace{2ex}

Consider the metric given in equation (\ref{ds}) where the
fields $(\lambda^0, \lambda^i, \lambda, \sigma)$ depend on $r$
only and obey the boundary conditions given in equation
(\ref{bc}) which correspond to asymptotically flat spacetime.
The ADM mass $M_{ADM}$ for such a metric is given, following the
analysis of \cite{jxlu}, by
\[
M_{ADM} = \frac{\omega_m \; V_{n_c}} {2 \kappa^2} \; \left(
m \; r^{m - 1} ( e^{2 \lambda} - \frac{e^{2 \sigma}}{r^2})
- r^m \partial_r ( m \; \frac{e^{2 \sigma}}{r^2} 
+ \sum_i e^{2 \lambda^i} ) \right)_{r \to \infty}
\]
where $\omega_m$ and $V_{n_c}$ are the volumes of the $m$
dimensional unit sphere and the $n_c$ dimensional torus. Let the
asymptotic behaviour of $(\lambda^0, \lambda^i, \lambda,
\sigma)$ in the limit $r \to \infty \;$ be given by
\[
\left( e^{2 \lambda^0}, \; e^{2 \lambda^i}, \; e^{2 \lambda}, \;
\frac{e^{2 \sigma}}{r^2} \right) \; \to \; 
1 - (c^0, \; c^i, \; c^r, \; c^\Omega) \; 
\frac{r_0^{m - 1}} {r^{m - 1}} 
\]
where $(c^0, \; c^i, \; c^r, \; c^\Omega)$ and $r_0$ are
constants. Evaluating the above expression for the ADM mass then
gives
\begin{equation}\label{ADM}
M_{ADM} = \frac{\omega_m \; V_{n_c} } {2 \kappa^2} \; 
\left( r_0^{m - 1} \; C_M \right)
\end{equation}
where 
\begin{equation}\label{CM}
C_M = m (c^\Omega - c^r) - (m - 1) \; (m c^\Omega + \sum_i c^i )
\; \; .
\end{equation}

We now write down the constants $(c^0, \; c^i, \; c^r, \;
c^\Omega)$ and $C_M$ for vacuum solutions, for $p-$brane
solutions , and for intersecting brane solutions.

\vspace{2ex}

\centerline{\bf Vacuum solutions} 

\vspace{2ex}

For vacuum solutions, we have
\[
\left( e^{2 \lambda^0}, \; e^{2 \lambda^i}, \; e^{2 \lambda}, \;
\frac{e^{2 \sigma}}{r^2} \right) = \left( e^{2 b^0 F}, \; 
e^{2 b^i F}, \; \frac{e^{2 b^\Omega F}}{f} , \; 
e^{2 b^\Omega F} \right) \; \; .
\]
From the behaviour of the solutions described in section {\bf 7}
in the asymptotic limit $\tau \to \tau_0$, it follows that
\[
f \; \to \; 1 - \frac{r_0^{m - 1}} {r^{m - 1}} 
\; \; , \; \; \; 
e^F \; \to \; 1 - \frac{r_0^{m - 1}} {r^{m - 1}} 
\]
in the limit $r \to \infty \;$. Hence, $(c^0, \; c^i, \;
c^\Omega) = 2 \; (b^0, b^i, b^\Omega)$ and $c^r = (2 b^\Omega -
1) \;$. It then follows from equation (\ref{CM}) that
\begin{equation}\label{CMvac}
C_{M \; (vac)} = m - 2 \; (m - 1) \; (m b^\Omega + \sum_i b^i)
\end{equation}
for vacuum solutions. In the standard case, we have
\[
2 \; (b^0, \; b^i, \; b^\Omega) = (1, \; 0, \; 0) 
\; \; , \; \; \; C_{M \; (vac)} = m \; \; .
\]


\vspace{2ex}

\centerline{\bf  $p-$branes} 

\vspace{2ex}

For M (string) theory $p-$brane solutions, the metric (in
Einstein frame) is described by
\[
\left( e^{2 \lambda^0}, \; e^{2 \lambda^i}, \; e^{2 \lambda}, \;
\frac{e^{2 \sigma}}{r^2} \right) = \left( H^{A^\parallel} \;
e^{2 b^0 F}, \; H^{A^\parallel \; or \; A^\perp} \; e^{2 b^i F},
\; H^{A^\perp} \; \frac{e^{2 b^\Omega F}}{f} , \; H^{A^\perp} \;
e^{2 b^\Omega F} \right)
\]
where $H = 1 + (1 - e^{2 w F}) \; {\cal S}^2$ with $w = b^0 +
\sum_{i \in \parallel} b^i$ and ${\cal S} = Sinh \; \Theta$, and
the exponents $A^\parallel$ is for directions parallel to the
brane and $A^\perp$ is for transverse directions. See equation
(\ref{dsr}) with ${\cal N} = 1$ and with $n_c \ge p$ in general.
In the limit $r \to \infty \;$, the functions $f$ and $e^F$ are
as given earlier, and hence
\[
H \to 1 + (2 \; w \; {\cal S}^2) \; 
\frac{r_0^{m - 1}} {r^{m - 1}} \; \; .
\]
Thus $ \left( e^{2 \lambda^0}, \; e^{2 \lambda^i}, \; e^{2
\lambda}, \; \frac{e^{2 \sigma}}{r^2} \right) \to 1 - (c^0, \;
c^i, \; c^r, \; c^\Omega) \; \frac{r_0^{m - 1}} {r^{m - 1}} \;$
in the limit $r \to \infty \;$ where, for a $p-$brane along
$(x^1, \cdots, x^p)$ directions,
\begin{eqnarray*} 
c^0 & = & 2 b^0 - 2 w A^\parallel \; {\cal S}^2  \\ 
c^i & = & 2 b^i - 2 w A^\parallel \; {\cal S}^2 
\; \; \; \; \; for \; \; \; \; i = 1, \cdots, p \\
& = & 2 b^i - 2 w A^\perp \; {\cal S}^2 
\; \; \; \; for \; \; \; \; i = p + 1, \cdots, n_c \\
c^\Omega & = & 2 b^\Omega - 2 w  A^\perp \; {\cal S}^2 \\
c^r & = & 2 b^\Omega - 1 - 2 w A^\perp \; {\cal S}^2 \; \; .
\end{eqnarray*}
Noting that $c^\Omega - c^r = 1$, and defining $X_H = p
A^\parallel + (m + n_c - p) A^\perp \;$, it then follows from
equation (\ref{CM}) that
\[
C_{M \; (p)} = m - 2 \; (m - 1) \; 
(m b^\Omega + \sum_i b^i - w X_H \; {\cal S}^2) \; \; . 
\] 
For M theory branes, $m + n_c = 9$ and $(A^\parallel, \;
A^\perp) = \left( - \frac{2}{3}, \; \frac {1} {3} \right)$ or
$\left( - \frac{1}{3}, \; \frac{2}{3} \right)$ for M2 or M5
branes. (For string theory $p-$branes $m + n_c = 8$ and it can
be shown that $ (A^\parallel, \; A^\perp) = \left( \frac{p -
7}{8}, \; \frac{p + 1}{8} \right)$ in Einstein frame.) In all
these cases, $A^\parallel - A^\perp = - 1$ and $(m + n_c) \;
A^\perp = p + 1 \;$. Hence, $X_H = p (A^\parallel - A^\perp) +
(m + n_c) A^\perp = 1$, and we have
\begin{equation}\label{CMp}
C_{M \; (p)} = m - 2 \; (m - 1) \; 
(m b^\Omega + \sum_i b^i - w \; {\cal S}^2) 
\end{equation}
for $p-$brane solutions. In the standard case, we have
\[
2 \; (b^0, \; b^i, \; b^\Omega \; ; \; w) = 
(1, \; 0, \; 0 \; ; \; 1)
\; \; , \; \; \; C_{M \; (p)} = m + (m - 1) {\cal S}^2 \; \; .
\]

\vspace{2ex}

\centerline{\bf Intersecting branes} 

\vspace{2ex}

For ${\cal N}$ sets of branes intersecting according to the BPS
rules, the $p-$brane analysis given above applies with only a
few straightforward modifications. Performing the analysis, it
can be seen that the final expression may be obtained upon
replacing $w {\cal S}^2$ in the $p-$brane expressions by $\sum_I
w^I {\cal S}_I^2 \;$ where $I = 1, 2, \cdots, {\cal N}$ and
$w^I$ are given in equation (\ref{wI}). Thus, 
\begin{equation}\label{CMNp}
C_{M \; ({\cal N} p)} = m - 2 \; (m - 1) \; (m b^\Omega 
+ \sum_i b^i - \sum_I w^I \; {\cal S}_I^2) 
\end{equation}
for intersecting brane solutions. In the standard case, we have
\[
2 \; (b^0, \; b^i, \; b^\Omega \; ; \; w^I) = 
(1, \; 0, \; 0 \; ; \; 1)
\; \; , \; \; \; C_{M \; ({\cal N} p)} = 
m + (m - 1) \sum_I {\cal S}_I^2 \; \; .
\]


\vspace{4ex}

\centerline{\bf Appendix C : Anayltical solutions for the 
functions $F$ and $f$ }

\vspace{2ex}

The equations of motion for $F$ and $f$ in terms of $r$ may be
obtained by substituting $\lambda^\alpha$ given in equation
(\ref{vaclambdaFr}) into vacuum equations of motion
(\ref{Lambdar2}) and (\ref{alpharr}) with $\Pi_I = p_{\alpha I}
= 0 \;$; or, by substituting $\lambda^\alpha$ given in equation
(\ref{lambdaFr}) into brane equations of motion (\ref{Pir}) --
(\ref{alpharr}). After some algebra, the resulting equations for
$F(r)$ and $f(r)$ can be written as
\begin{eqnarray}
2 (B - b^\Omega) \; f \; (r F_r) & = & 
(m - 1) (1 - f) + \frac{K}{m} \; f \; (r F_r)^2 \label{e1} \\
2 (B - b^\Omega) \; f \; (r F_r) & = & 
2 (m - 1) (1 - f) - r f_r \label{e2} \\
e^{(B - b^\Omega) F} \; \sqrt{f} \; (r F_r) 
& = & \frac{{\cal M}}{r^{m - 1}} \label{e3}
\end{eqnarray}
where ${\cal M} = (m - 1) \; r_0^{m - 1}$ and $B$ and $K$ are
defined in equation (\ref{BK}). Equation (\ref{e3}) follows
directly from equation (\ref{F}), namely from $e^{\Lambda -
\lambda} \; F_r = {\cal M} \;$. We now set $2 (B - b^\Omega) =
1$ with no loss of generality. For $K = 0$, the above equations
can be solved easily and the solutions are given by
\[
f = e^F = 1 - \frac{r_0^{m - 1}}{r^{m - 1}} \; \; . 
\]

It turns out that equations (\ref{e1}) -- (\ref{e3}) can be
solved analytically for $K > 0$ also; and, that the properties
of the functions $F(r)$ and $f(r)$ can be understood even
without obtaining solutions. These properties and the explicit
analytical forms of the functions $F$ and $f$ have been
described in our earlier reports \cite{k11}. For the sake of
completeness, we present these solutions here along with a
description of their derivation and some of their properties.

Define $R = r^{m - 1}$ and $R_0 = r_0^{m - 1} \;$. We then get 
\begin{eqnarray*}
f \; (R F_R) & = & 1 - f + \frac{(m - 1) K}{m} \; f \; (R F_R)^2
\label{e1A} \\
f \; (R F_R) & = & 2 (1 - f) - R f_R \label{e2A} \\
e^F \; f \; (R F_R)^2 & = & \frac{{R_0^2}}{R^2} \label{e3A}
\end{eqnarray*}
where the subscripts $R$ denote $R-$derivatives. It follows that
\[
(1 - f) - R f_R = \frac{(m - 1) K}{m} \; f \; (R F_R)^2 =
\frac{(m - 1) K}{m} \; 
\frac { \left(2 (1 - f) - R f_R \right)^2} {f}
\]
which gives a quadratic equation for $R f_R \;$, and that 
\[
e^F = \frac {R_0^2 \; f} 
{R^2 \; \left( 2 (1 - f) - R f_R \right)^2} \; \; .
\]
Defining 
\[
b = \frac{4 (m - 1) K}{m} \; \; , \; \; \; 
\alpha = \frac{1}{1 + b} \; \; , \; \; \; 
f_0 = 1 - \alpha 
\]
and solving the quadratic equation for $R f_R$ gives, after some
algebra,
\begin{equation}\label{RfR} 
R f_R = \frac{2 \; (1 - f) \; (f - f_0)} 
{f - f_0 + \epsilon \; \sqrt{\alpha \; f \; (f - f_0)}}
\end{equation} 
and 
\begin{equation}\label{eF} 
e^F = \frac{R_0^2 \; \left( \sqrt{f - f_0} 
+ \epsilon \; \sqrt{\alpha \; f} \right)^2 } 
{4 \; \alpha \; R^2 \; (1 - f)^2}
\end{equation} 
where $\epsilon = \pm 1$ and the square roots are always to be
taken with a postive sign.

We consider the case $K > 0$, hence $b > 0$, so that $\alpha <
1$ and $f_0 > 0 \;$. Define $R_{min}$ by $f(R_{min}) = f_0 \;$,
and a function $g(R)$ by
\begin{equation}\label{hR} 
\sqrt{f - f_0} = \epsilon_g \sqrt{\alpha} \; g 
\; \; \; \Longrightarrow \; \; \; f = 1 - \alpha + \alpha g^2
\end{equation}
where $\epsilon_g = Sgn \; g$ and, further, choose $\epsilon_g =
\epsilon$ with no loss of generality. Using the asymptotic
behaviour of $f$, namely $f \to 1 - \frac{R_0}{R}$ as $R \to
\infty \;$, it can be seen from equation (\ref{RfR}) that
$\epsilon = + 1 \;$, and from equation (\ref{hR}) that $g(R) \to
1 - \frac {R_0} {2 \alpha R} \;$, in this limit. It can further
be seen that $\; g > 0$ and $\epsilon_g = + 1$ for $R_{min} < R
< \infty \;$; that $g(R_{min}) = 0 \;$; and that $g < 0$ and
$\epsilon_g = - 1 \;$ for $0 < R < R_{min} \;$. Writing $f$ in
terms of $g$ and after some algebra, equations (\ref{RfR}) and
(\ref{eF}) now become
\begin{equation}\label{RhR}
R g_R = \frac{1 - g^2}{g + \sqrt{f}}
\; \; \; \; \Longrightarrow \; \; \; \; 
\frac{d R}{R} = d g \left( \frac{g + \sqrt{f}}{1 - g^2} \right)
\end{equation}
and 
\begin{equation}\label{e^F}
e^F = \frac{R_0^2}{4 \alpha^2 R^2} \; 
\left( \frac{g + \sqrt{f}} {1 - g^2} \right)^2 \; \; .
\end{equation}
It turns out that equation (\ref{RhR}) can be solved and an
explicit analytical solution for $R$ in terms of $g$ can be
obtained, because
\[
\frac{g + \sqrt{f}}{1 - g^2} 
= \frac{g}{1 - g^2} - \frac{\alpha}{\sqrt{f}} 
+ \frac{1}{2} \; \left( \frac{1}{1 - g} 
+ \frac{1}{1 + g} \right) \; \frac{1}{\sqrt{f}} 
\]
and each term on the right hand side can be integrated in a
closed analytical form. Incorporating $g(R_{min}) = 0 \;$,
equivalently $R(0) = R_{min} \;$, we get
\begin{eqnarray*}
ln \; \frac{R_{min}}{R} & = & \frac{1}{2} \; ln \; 
\vert 1 - g^2 \vert
+ \sqrt{\alpha} \; Sinh^{- 1} \; \frac{g \; \sqrt{\alpha}}
{\sqrt{1 - \alpha}} \\
& & - \frac{1}{2} \; Sinh^{- 1} \; 
\frac{1 - \alpha + \alpha g}{(1 - g) \; \sqrt{\alpha (1 - \alpha)}} 
\\ & &
+ \frac{1}{2} \; Sinh^{- 1} \; 
\frac{1 - \alpha - \alpha g}{(1 + g) \; \sqrt{\alpha (1 - \alpha)}} 
\; \; .
\end{eqnarray*}
A more explicit expression for $R(g)$ can be obtained, and in
many equivalent forms, by using identities such as
\[
Sinh^{- 1} \; x = ln \; \left( x + \sqrt{1 + x^2} \right) 
\]
\[
f - g^2 = \left( 1 - \alpha \right) \left( 1 - g^2 \right)
\]
\[
\alpha \left( 1 - \alpha \right) \left( 1 \pm g \right)^2 
+ \left( 1 - \alpha \mp \alpha g \right)^2 = f
\]
\[
\left( g - \sqrt{f} \right) \; 
\left( 1 - \alpha - \alpha g - \sqrt{f} \right) = 
\left( 1 - \alpha \right) \left( 1 + g \right) \left( 1 - \sqrt{f}
\right)
\]
\[
\left( 1 - \alpha - \alpha g + \sqrt{f} \right) \; 
\left( 1 - \alpha + \alpha g + \sqrt{f} \right) 
= \left( 1 - \alpha \right) \; \left( 1 + \sqrt{f} \right)^2
\; \; . 
\]
The expression we find convenient is given, after a series of
manipulations, by
\begin{equation}\label{r(h)}
\frac{R_{min}}{R} = 
\frac{\sqrt{1 - \alpha} \; \; (1 - g) \; (1 + \sqrt{f})} 
{1 - \alpha + \alpha g + \sqrt{f}} \; \; 
\left( \frac{\sqrt{f} + g \sqrt{\alpha}}{\sqrt{1 - \alpha}}
\right)^{\sqrt{\alpha}} \; \; . 
\end{equation}
Equation (\ref{e^F}) now gives an expression for $e^F$ in terms
of $g \;$.

\vspace{2ex} 

We now describe some properties of the analytical solutions
given above. These properties all follow straightforwardly, but
after some algebra.

\begin{itemize}

\item

The above expression for $R(g)$ satisfies the differential
equation (\ref{RhR}).

\item

Consider $\alpha = 1 \;$. Noting that $\left( \sqrt{1 - \alpha}
\right)^{1 - \sqrt{\alpha}} = 1$ in the limit $\alpha \to 1 \;$,
and taking $\sqrt{f} = g \;$, we get $\frac{R_0}{R} = 1 - g^2 =
1 - f \;$. It then follows from equations (\ref{RhR}) and
(\ref{e^F}) that $e^F = f = 1 - \frac{R_0}{R} \;$.

\item

Consider $\alpha < 1 \;$. As $R$ decreases from $\infty$ to
$R_{min}$ to $R_1$ to $0 \;$, $\; g$ decreases from $1$ to $0$
to $- 1$ to $- \infty \;$; hence, $f$ decreases from $1$,
reaches its minimum $f_0$, then increases and reaches $1$ again,
and increases further to $\infty \;$. Thus, $R(g)$ for some
select values of $g$ are:
\[
R(1_-) = \infty \; \; , \; \; \; 
R(0) = R_{min} \; \; , \; \; \; 
R(- 1) = R_1 \; \; , \; \; \; 
R(- \infty) = 0 \; \; . 
\]

\item

As $R$ decreases from $\infty$ to $0 \;$, $\; e^F$ decreases
monotonically from $1$ to $0$, remaining $< f$ always.

\item 

Consider the limit $R \to \infty \;$.  Setting $1 - g =
\frac{R_0}{2 \alpha R} \;$ and $f = g = 1 \;$ in equation
(\ref{r(h)}) gives
\[
R_{min} = \frac{c(\alpha)}{\alpha} \; R_0
\; \; \; , \; \; \; \; 
c(\alpha) = \frac{1}{2} \; \left( 1 + \sqrt{\alpha} 
\right)^{\frac{1 + \sqrt{\alpha}}{2}}
\; \left( 1 - \sqrt{\alpha} 
\right)^{\frac{1 - \sqrt{\alpha}}{2}} \; \; , 
\]
which thus expresses $R_{min}$, where the function $f$ reaches
its minimum, in terms of $R_0 = r_0^{m - 1} \;$. Note that
$\frac{1}{2} \le c(\alpha) \le 1$ for $0 \le \alpha \le 1$, as
can be shown easily.

\item 

Consider $R = R_1 \;$. Then $g = - 1$ and $f = 1 \;$. Equation
(\ref{r(h)}) now gives
\[
R_1 = c(\alpha) \; R_{min} \; \; .
\]
which thus expresses $R_1$ in terms of $R_{min}$ and thereby in
terms of $R_0 \;$.

\item 

Consider the limit $R \to 0 \;$. Then $g \to - \infty \;$,
$\sqrt{f} \to (- g) \sqrt{\alpha} \;$, and $ \sqrt{f} + g
\sqrt{\alpha} = \frac{f - \alpha g^2}{\sqrt{f} - g
\sqrt{\alpha}} \; \to \; \frac{1 - \alpha}{2 \sqrt{\alpha}} \;
\frac{1}{(- g)} \;$. We then get
\[
R^{- 1} \sim (- g)^{1 - \sqrt{\alpha}} \; \; , \; \; \;
f \sim g^2 \sim R^{- \frac{2}{1 - \sqrt{\alpha}}} \; \; .
\]

\item 

In the main body of the paper, the general solutions for $R =
r^{m - 1}$, $\; F$, and $f$ have been obtained in terms of $\tau
\;$, and are given in equations (\ref{r(tau)vac}) and
(\ref{sqrtfvac}). Using equation (\ref{hR}), one can now obtain
an expression for $g$ also in terms of $\tau \;$. It can then be
shown, after a long but straightforward algebra, that these
expressions for $R(\tau)$, $\; F(\tau)$, $\; f(\tau)$, and
$g(\tau)$ satisfy equations (\ref{e^F}) and (\ref{r(h)}).

\end{itemize}




\end{document}